\documentclass[twocolumn,longbibliography,floatfix,tightline]{revtex4-1}
\usepackage{amssymb}
\usepackage{amsmath}
\usepackage{array}
\usepackage{colordvi}
\usepackage[colorlinks]{hyperref}
\usepackage{amsthm}
\usepackage{subeqnarray}
\usepackage{cases}
\usepackage{enumerate}
\usepackage{graphicx}
\usepackage{epsfig}
\usepackage{epstopdf}
\usepackage{subfigure}
\usepackage{color}
\usepackage{bm}
\usepackage[switch,displaymath]{lineno}

\def\avg(#1){\langle#1\rangle}

\def\be{\begin{equation}}
\def\ee{\end{equation}}
\def\bea{\begin{eqnarray}}
\def\eea{\end{eqnarray}}

\def\nn{\nonumber}

\usepackage{ulem}

\begin{document}

%\preprint{APS/123-QED}

\title{Topological spinor vortex matter on spherical surface induced by non-Abelian spin-orbital-angular-momentum coupling}% Force line breaks with \\
\author{Jia-Ming Cheng, Ming Gong, Guang-Can Guo, Zheng-Wei Zhou,}
\author{Xiang-Fa Zhou}
\email{xfzhou@ustc.edu.cn}
\affiliation{CAS Key Laboratory of Quantum Information, University of Science and Technology of China
, Hefei, 230026, China \\
CAS Center For Excellence in Quantum Information and Quantum Physics, University of Science and Technology of China, Hefei, 230026, China }

\date{\today}% It is always \today, today,

\begin{abstract}
We provide an explicit way to implement non-Abelian spin-orbital-angular-momentum (SOAM) coupling in spinor Bose-Einstein condensates using magnetic gradient coupling.
For a spherical surface trap addressable using high-order Hermite-Gaussian beams, we show that this system supports various degenerate ground states carrying different total angular momenta $\mathbf{J}$, and the degeneracy can be tuned by changing the strength of SOAM coupling.
For weakly interacting spinor condensates with $f=1$, 
the system supports various meta-ferromagnetic phases and meta-polar states described by quantized total mean angular momentum $|\avg(\mathbf{J})|$.
Polar states with $Z_2$ symmetry and Thomson lattices formed by defects of spin vortices are also discussed.
The system can be used to prepare various stable spin vortex states with nontrivial topology, and serve as a platform to investigate strong-correlated physics of neutral atoms with tunable ground-state degeneracy.
\end{abstract}

%\pacs{03.75.Lm, 74.50.+r, 32.10.Fn}% PACS, the Physics and Astronomy
\maketitle

%\tableofcontents

%\section{introduction}
\section{Introduction}
Spin-orbital-angular-momentum (SOAM) coupling, originally introduced due to the relativistic effect of the electron's spin with its orbital angular momentum, is of ubiquitousness now in varying areas of physics. %\cite{galitski2013spin,zhai2015degenerate,zhou2013unconventional,manchon2015new,aleiner2001spin,fedorov2003spin,bliokh2015spin}.
%Especially in condensed matter physics, SOAM coupling plays a vital role for many novel phenomena such as spin Hall effect \cite{sinova2015spin},
%topological insulator and superconductivity \cite{hasan2010colloquium,qi2011topological}, etc..
For neutral atoms, recent investigations show that the internal atomic spin do can be coupled to its momentum degree of freedom with the help of laser beams 
\cite{jaksch2003creation,zhu2007simulation,lin2011spin,zhang2012collective,zhang2013topological,ji2014experimental,wu2016realization,huang2016experimental}.
%Experimentally the technique works for both bosons and fermions with much flexibility \cite{}.
%This enables us to study such novel effects in a highly controllable manner.
Since the first experimental realization of spin-momentum coupling (SMC) in the condensate of $^{87}\text{Rb}$ atoms\cite{lin2011spin}, various experimental and theoretic efforts have been made along this direction \cite{galitski2013spin,zhai2015degenerate,zhou2013unconventional,wang2010spin,cong2011unconventional,
hu2012spin,li2013superstripes,ozawa2012stability,lobanov2014fundamental,zhang2015stable,luo2017spin,
li2017stripe,wu2016realization,huang2016experimental,han2018chiral,wu2018beliaev,liao2018searching,
qu2018angular,qu2013topological,hu2011probing,wu2013unconventional,sinha2011trapped,li2016three,cole2012bose,cheng2018symmetry}.

%For instance, the celebrated standing-wave phase has been observed in \cite{li2017stripe}, and experimental implementing 2D SMC is also under active investigations \cite{wu2016realization,huang2016experimental}.
%Theoretically, the presence of SMC in ultra-cold gases also induces various exotic quantum phases including
%topological superfluidity \cite{zhang2013topological,qu2013topological,hu2011probing,wu2013unconventional},
%novel spin textures \cite{sinha2011trapped,cong2011unconventional,hu2012spin,zhou2013unconventional,li2016three,cole2012bose}, plane waves phases \cite{wang2010spin,cheng2018symmetry}, etc..

Most current investigations focus on the spin-momentum coupled systems.
For the usual SOAM interaction, theoretical and experimental investigations are considered recently only for the Abelian type interaction $L_zF_z$  \cite{marzlin1997vortex,liu2007optically,sun2015spin,qu2015quantum,hu2015half,demarco2015angular,chen2018spin,PhysRevLett.122.110402,chen2018rotating,chen2019angular}.
However, the original SOAM coupling $\mathbf{L}\cdot \mathbf{F}$ in atomic physics is non-Abelian.
Such symmetric non-Abelian feature results in various fine structures of atomic levels with different degeneracy \cite{cowan1981theory}.
Meanwhile, this interaction is also closely linked with the generalization of quantum Hall physics in 3D system \cite{li2013topological,li2013high}, and can be viewed as the parent system to generate almost all relevant spin-orbital interactions discussed in current studies \cite{li2012two,li2013high}.
However, the realization of such non-Abelian SOAM coupling seems extremely difficult which greatly constrains our abilities to explore such novel physics in cold atoms.

On the other hand, there is also a growing interest in the effect of the underlying geometry on various quantum orders \cite{ho2015spinor,sun2018static,zhang2018potential,tononi2019bose,batle2016generalized,fomin2012tunable,PhysRevLett.114.197204,parente2011spin,li2015topological,imura2012spherical,kraus2008testing,moroz2016chiral,shi2015emergent}.
For cold atoms, exotic vortex structures on a cylindrical surface have been considered recently by Ho and Huang \cite{ho2015spinor}.
Meanwhile, spherical shell geometry induced by hedge-hog like gradient magnetic fields has also be proposed for spinful atoms with larger internal spin $F$ \citep{zhou2018synthetic}.
In all these constructions, the atomic spin is frozen along the external magnetic fields, which inhibits the investigation of SOAM coupled physics in curved geometry.  
The construction of a perfect spherical surface trap with magnetic polarization for cold atoms also remains as an another experimental urgent task.

In this paper, with the help of a time-dependent hedge-hog-type gradient magnetic fields (which proves to be possible for spinful atoms)\citep{zhou2018synthetic,goldman2014periodically}, we show that non-Abelian SOAM coupling do can be implemented in cold atomic systems.
We further show that by constructing an effective curved surface trap using high-order Hermite-Gaussian laser beams, we can change the SOAM coupling strength in a wide range of parameters.
Thanks to the high symmetry of the system, the system supports ground-states with tunable degeneracy.

For weakly interacting spinor condensates with $f=1$ \citep{ho1998spinor,kawaguchi2012spinor}, the system support various meta-ferromagnetic (mFM) and meta-polar (mP) phases with quantized magnitudes of the total angular momentum (TAM) $|\avg(\mathbf{J})|=|\avg(\mathbf{L}+\mathbf{F})|$ and non-vanishing spin fluctuations.
This is completely different from the usual vortex phases characterized by the quantized angular momentum $L_z$ only.
In the case of vanishing spin-exchange interaction, the defects of spin vortices form stable lattice configurations on sphere characterised by the standard Thomson problem \citep{thomson1904xxiv,zhou2018synthetic}. 
Meanwhile, in the polar regime with strong spin-exchange interaction, the system supports stable nontrivial polar states \citep{kawaguchi2012spinor,zibold2016spin,lovegrove2016stability} characterized by $Z_2$-type topological invariant.
The system can be viewed as a vortex zoo of constructing stable spin vortices with novel intrinsic topology, and serve as a desired platform to explore various non-Abelian SOAM coupled physics for both atomic species.
%, and also provides a route to investigate various strong-correlated physics on a curved spherical geometry.

\section{Scheme of implementing SOAM coupling and the model Hamiltonian}
To illustrate the novel physics induced by such spherical surface trap, we consider spinor condensates suffering from an isotropic non-Abelian SOAM coupling described by $\mathbf{L}\cdot\mathbf{F}$.
Such 3D SOAM coupling occurs in atomic physics due to the relativistic effect, where the spin of electrons only take values $S=1/2$.
While for neutral atoms, $F$ can take integer and half-integer values for boson and fermions, which greatly enriches the underlying physics. 
However, the implementation of such non-Abelian coupling is nontrivial for cold atoms as the relativstic effect is almost undetectable. 

To implement the isotropic SOAM coupling $\mathbf{L}\cdot\mathbf{F}$ for atoms, we introduce a Zeeman coupling term $\mathbf{r}\cdot\mathbf{F}$ involving an effective hedge-hog type magnetic gradient fields.
This monopole-like effective magnetic field has recently be proved to be possible for atoms with internal spin $\mathbf{F}$ \cite{zhou2018synthetic} by employing the following Floquet engineered quadrapole field
\begin{align}
		\mathbf{B}=B_0\mathbf{e}_z+B_1[1-4\cos(\omega_0 t)](x\mathbf{e}_x+y\mathbf{e}_y-2z\mathbf{e}_z)
\end{align}
with a strong bias field $B_0$. The single-particle Hamiltonian can be written as
\begin{align}
		H^{\prime}=H_0-\mu_Bg_F\mathbf{B}\cdot\mathbf{F} \mbox{ with }H_0=-\frac{\hbar^2\mathbf{\nabla}^2}{2\mu}+V(r). \nn
\end{align}	
In current experiments, the frequency $\omega_0$ can be set up to $10^6$ Hz. 
If we choose $B_0$ such that $\omega_0=\Omega=\mu_Bg_FB_0/\hbar$, and assume that this frequency is much larger than all the other energy scales, then in the rotating frame defined by $U=e^{-i\Omega tF_z}$, the effective Hamiltonian reads
\begin{align}
	H=U^{\dagger}H^{\prime}U-i\hbar U^{\dagger}\partial_tU=H_0-2\mu_Bg_FB_1\mathbf{r}\cdot\mathbf{F},
\end{align}
where an effective monopole-like magnetic fields is induced.
%When $B_0=6.0$ G, frequency $\omega_0=\Omega\approx2\pi\times 4.2$ MHz.	For atoms $^{87}$Rb, $g_F=-1/2$.

The implementation of 3D non-Abelian SOAM coupling is apparent now if we consider the following time-dependent Hamiltonian with a hedge-hog-type magnetic Zeeman coupling in the rotating frame \cite{zhou2018synthetic,goldman2014periodically}
\bea
H(t)=H_0+v_0\cos(\Omega_0t) \mathbf{r}\cdot\mathbf{F}  \label{Ht_rF}
\eea
with $v_0=-2\mu_Bg_FB_1$. 
Here, the driven frequency $\Omega_0$ (about $10^4 \sim 10^5$ kHz) should be choosen such that $\omega_0 \gg \Omega_0 \gg \omega$ is satisfied. In this case, using the commutation relation
$[\mathbf{r}\cdot\mathbf{F},\mathbf{p}\cdot\mathbf{F}]=i\hbar(\mathbf{F}^2+\mathbf{L}\cdot\mathbf{F})$,
the dynamics of the system can be described by the following effective interaction
\begin{align}
	H_{\text{eff}} = H_0+\tilde{\lambda}(\mathbf{F}^2+\mathbf{L}\cdot\mathbf{F})+O \left(\frac{1}{\Omega^3_0} \right),
\end{align}
where the SOAM coupling strength reads $\tilde{\lambda}=v_0^2/(\mu\hbar^2\Omega^2_0)$.
Thus, up to a constant term $\mathbf{F}^2$, we have succeeded in realizing the desired ($\mathbf{L}\cdot\mathbf{F}$)-type coupling.
Although $\tilde{\lambda}$ is always positive in the above case, we note that negative SOAM coupling can also be implemented using a two-step scheme with modified magnetic gradient pulses.
%Specifically, in the first step, a traditional 3D weyl-type SOC $\mathbf{P}\cdot\mathbf{F}$ is realized with the help of the linear gradient magnetic pulses composed of $r_iF_i(i=x,y,z)$; and then an effective hedge-hog-type gradient magnetic field is employed to implement the desired negative SOAM coupling. 
Explicit construction of these pulse sequences can be found in Appendix A.

%\begin{figure}[htbp]
%	\includegraphics[width=0.45\textwidth]{./figure/spherical_trap.eps}% Here is how to import EPS art
%	\caption{\label{fig:sphe_trap1}
%	Spherical trap in case of $U=8$, $U_0=200$, $w_1=10$.}
%\end{figure}
%%%%%%%%%%%%%%%%%%%%%%%%%%%%%%%%%%%%%%%%%%%%%%%%%%%%%%%%

We stress that the above expansion works only when $v_0/\Omega_0 \ll 1$ is satisfied.
This greatly reduces the range of the parameter as $\tilde{\lambda} \ll 1$, which makes the non-Abelian feature of the system hard to observe.
However, as will be shown latter, the presence of a curved spherical surface trap in Eq.(\ref{eq:sphere})  can greatly enhance the effective coupling.
Such spin-independent trap liberates the spin degrees of freedom, and enables the investigation of various phases with novel spin textures induced by non-Abelian SOAM coupling.

%Physically, such hollow symmetric spherical surface trap cannot be simply implemented using the Gaussian beam TEM$_{00}$ only.
%To overcome this difficulty, here we propose to introduce the combined laser beams composed of high-order Hermite-Gaussian beams TE$_{mn}$, where the electric field amplitude reads (see SM\citep{SM} for details)
%\begin{align}
%	E_{mn}(x,y,z)\propto H_m(\frac{\sqrt{2}x}{w_1})H_n(\frac{\sqrt{2}y}{w_1})e^{-\frac{x^2+y^2}{w^2}}e^{ikz},
%\end{align}
%with $w_1$ the waist of the beam and the first few Hermite polynomials
%\bea
%H_0(x)=1, \hspace{.5cm} H_1(x)=2x, \hspace{.5cm} H_2(x)=4x^2-2. \nn 
%\eea
%For beams traveling along the $z$-axis, appropriate combinations of TEM$_{20}$, TEM$_{11}$, and TEM$_{00}$ modes can result in an effective potential $\propto x^4+2x^2y^2-\alpha(x^2+y^2)$ with tunable parameter $\alpha$.
%When these beams are employed along the other two directions as indicated in Fig. \ref{fig:sphe_trap}, we can construct an ideal spherical surface trap $V(r)$ with both tunable radius $R$ and trapping frequency $\omega$.

\section{Spherical surface trap created using high-order Hermite-Gaussian beams}\label{app:sph_trap}
To construct a perfect curved surface traps for spinor atoms, we propose to employ high-order Hermite-Gaussian laser beams, with which the spin degree of freedom of the atoms can be liberated.
For usual Hermite-Gaussian mode denoted as TEM$_{mn}$, the electric field amplitude reads
\begin{align}
	E_{mn}(x,y,z)\propto H_m(\frac{\sqrt{2}x}{w_1})H_n(\frac{\sqrt{2}y}{w_1})e^{-\frac{x^2+y^2}{w_1^2}}e^{ikz},
\end{align}
which is propagating along $z$ direction with the waist radius $w_1$.
Here $H_m(x)$ is the Hermite polynomial.
The first few polynomials are listed as follows for later convenience
\bea
&H_0(x)=1, \hspace{.25cm} H_1(x)=2x, \hspace{.25cm} H_2(x)=4x^2-2,  \cdots.
\eea

%%%%%%%%%%%%%%%%%%%%%%%%%%%%%%%%%%%%%%%%%%%%%%%%%%%%%
\begin{figure}[htpb]
	\includegraphics[width=0.35\textwidth]{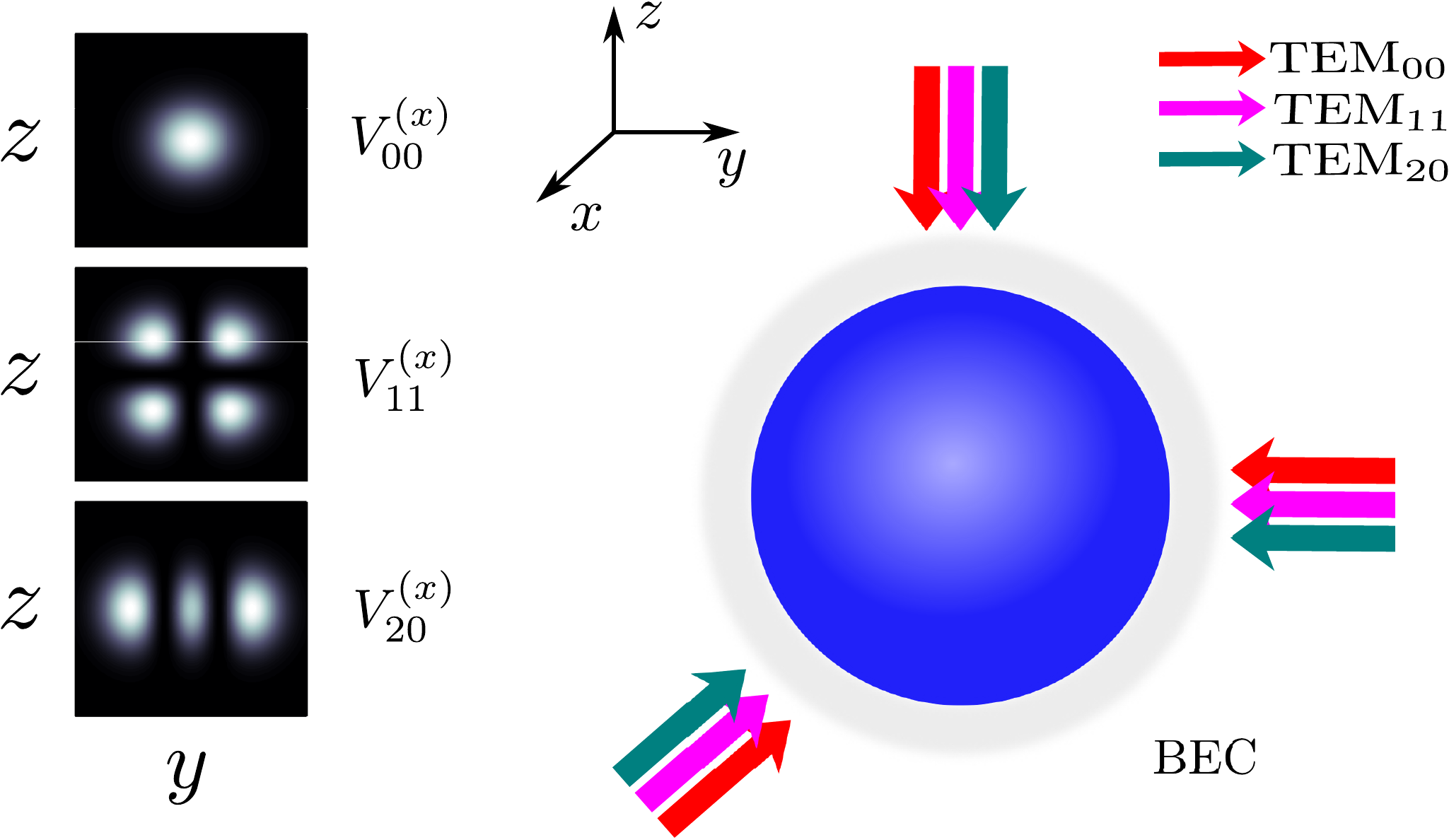}% Here is how to import EPS art
	\caption{\label{fig:sphe_trap}
	Schematic plot about the implementation of the spherical surface trap using Hermite-Gaussian modes along $x$, $y$, and $z$-axis respectively. The inserts show the sectional view of these modes.}
\end{figure}
%%%%%%%%%%%%%%%%%%%%%%%%%%%%%%%%%%

%%%%%%%%%%%%%%%%%%%%%%%%%%%%%%%%%%%%%%%%%%%%%%%%%%%%%%%%%
%\begin{figure}[htbp]
%	\includegraphics[width=0.48\textwidth]{./figs2.eps}% Here is how to import EPS art
%	\caption{\label{fig:sphe_trap1}
%	(a) Minimal points of Optical trap $V(x,y,z)$ forming a spherical surface and (b) optical trap $V(x,0,0)$ along $x$ direction in case of $U=8$, $U_0=200$, $w_1=10$.}
%\end{figure}
%%%%%%%%%%%%%%%%%%%%%%%%%%%%%%%%%%%%%%%%%%%%%%%%%%%%%%%%%%%

To obtain a spherical surface trap, we can employ composite Hermite-Gaussian modes along every direction.
For instance, in the case of the electric dipole approximation, three modes such as TEM$_{00}$, TEM$_{11}$, and modified TEM$_{20}$ (see Fig. 1) along the $z$-axis can induce the following optical potentials
\begin{align}
	V_{00}^{(z)}=&U_0e^{-\frac{2(x^2+y^2)}{w^2_1}}\approx U_0 \left[1-\frac{2(x^2+y^2)}{w^2_1} \right],\nn \\
	V_{11}^{(z)}=&U_1x^2y^2e^{-\frac{2(x^2+y^2)}{w^2_1}}\approx U_1x^2y^2\left[1-\frac{2(x^2+y^2)}{w^2_1} \right],\nn \\
	V_{20}^{(z)}=&U_2x^4e^{-\frac{2(x^2+y^2)}{w^2_1}}\approx U_2x^4\left[1-\frac{2(x^2+y^2)}{w^2_1} \right], \nn
%	V_{20}^{(z)}=&U_2(x^2-\frac{1}{2})^2e^{-\frac{2(x^2+y^2)}{w^2_1}} \nn \\
%	\approx& U_2(x^4-x^2+\frac{1}{4})\left[1-\frac{2(x^2+y^2)}{w^2_1} \right],
\end{align}
where we have used the paraxial approximation such that $w_1 \gg \{x,y,z\}$. The total potential after summing over all those along the $x$, $y$ and $z$ direction reads
\begin{align}
	V =&\sum_{i=x,y,z}(V_{00}^{(i)}+V_{11}^{(i)}+V_{20}^{(i)})\approx U_2(x^4+y^4+z^4)\notag \\
	& + U_1(y^2z^2+z^2x^2+x^2y^2)-\frac{4U_0}{w^2_1}r^2+3U_0,  \nn
%	&U_1(y^2z^2+z^2x^2+x^2y^2)-(U_2+\frac{4U_0}{w^2_1})r^2+3U_0+\frac{U_2}{4},  \nn
\end{align}
where $r^2=x^2+y^2+z^2$, and other terms proportional to $r^6$ or higher are safely neglected due to paraxial conditions. Then if we set $U_1=2U_2=2U$ and $R^2=2U_0/(Uw^2_1)$, %$R^2=1/2+2U_0/(Uw^2_1)$,
the total potential field can be approximated as $V\simeq U(r^2-R^2)^2-UR^4+3U_0$, 
which indicates that the minimal value of the total potential is obtained at $r=R$. Around this minimal point, total potential can be rewritten as (up to a constant term)
\begin{align}
	\label{eq:sphere}
	V(r)\approx 4UR^2(r-R)^2 = \frac{1}{2}\mu\omega^2(r-R)^2 
\end{align}
with $\mu$ the mass of the atom, and $\omega=\sqrt{8UR^2/\mu}$ the characteristic frequency.
We stress that to ensure the paraxial conditions, appropriate $U_0$, $U$ and $w_1$ should be chosen so that $R\ll w_1$ is satisfied.

Eq. (\ref{eq:sphere}) represents a perfect spherical surface trap with tunnable radius $R$ and trapping frequency $\omega$ induced by laser beams. 
This provides an ideal platform of investigating various novel physics for cold atoms subject to such boundaryless curved geometry. 
Especially, for SOAM coupled condensates, the system supports exotic spinor vortex phases with intrinsic topological properties. 

%For instance, if we set the radius of the spherical trap as the unit of length with $R \equiv 1$ and assume $w_1=10 \gg R$, then the trapping amplitude of different laser beams should satisfy $U_0=w_1^2R^2U/2=50U$.
%The characteristic length $l_T$ of the harmonic potential around the minimal of the total potential can then be adjusted by the amplitude $U$, which approximately equals $\sqrt{\hbar/(\mu\omega)}$. It is about $0.48~\mu$m in case of $\omega=2\pi\times 0.5$ kHz for $^{87}$Rb atoms.

%\begin{figure}[htbp]
%	\includegraphics[width=0.35\textwidth]{./figure/sphericaltrap.eps}% Here is how to import EPS art
%	\caption{\label{fig:sphe_trap}
%	Schematic plot about the implementation of the spherical surface trap using Hermitian-Gaussian modes along $x$, $y$, and $z$-axis respectively. The inserts show the sectional view of these modes.}
%\end{figure}

%\section{\label{sec:single}single particle spectra}
\section{Single particle spectra}

In the case of deep traps and low temperature, the radial motion of atoms is frozenthe and atoms are mainly confined around the spherical surface with radius $R$. 
The field operator can be assumed to be $\varphi(r)\psi(\theta,\phi)$.
The radial wavefunction $\varphi(r)$ reads $\varphi(r)=\left( \pi l_T^2r^4 \right)^{-1/4}\exp\left[-(r-R)^2/(2l_T^2)\right]$, where $l_T= \sqrt{\hbar/\mu\omega}\ll R$ is the characteristic length of spherical trap in radial direction.
After integrating out the radial degree-of-freedom, we obtain a reduced dimensionless single-particle Hamiltonian in a spherical surface trap as (See Appendix B)
\begin{align}
	\label{eq:H0}
	\mathcal{H}_0=\mathbf{L}^2+\lambda\mathbf{L}\cdot\mathbf{F},
\end{align}
where $\lambda=2\mu R^2\tilde{\lambda}=2\mu^2_Bg^2_FB^2_1R^2/(\hbar^2\Omega^2_0)$, which thus can be tuned in a wide range by changing the radius $R$, or the ratio $B_1$ respectively.

The system possesses conserved quantities including $\mathbf{L}^2$, $\mathbf{F}^2$, $\mathbf{J}^2$, and $J_z$ with the TAM $\mathbf{J}=\mathbf{L}+\mathbf{F}$.
The single-particle eigenstates can be labeled using quantum numbers $[l, f, j, j_z]$ as 
\bea
\psi^{l,f}_{j,j_z}(\theta,\phi)= \vert j, j_z\rangle=\sum_{m=-l}^{l}\sum_{f_z=-f}^{f} C^{(l,f)}_{j,j_z}\vert l,m;f,f_z\rangle
\eea
with the Clebsch-Gordan (CG) coefficients $C^{(l,f)}_{j,j_z} \equiv \langle l,m;f,f_z\vert j,j_z\rangle$, $\vert l,m\rangle\equiv Y_{l,m}(\theta,\phi)$ the usual spherical harmonics, and $\vert f,f_z\rangle$ the internal state of spinful atoms. The corresponding single-particle energy is degenerate for different $j_z$ and reads
\begin{align}
	\mathcal{E}_0=l(l+1)+\frac{\lambda}{2}[j(j+1)-l(l+1)-f(f+1)]
\end{align}
with $j=\vert l-f\vert,\vert l-f\vert+1,\cdots, l+f$.
When $\lambda>0$, $\mathbf{L}$ is anti-parallel to $\mathbf{F}$, and we have $j=\vert l-f\vert$ for the ground state.
Otherwise, we have $j=l+f$.
In both cases, the explicit values of $l$ and $j$ for ground states depend on the coupling $\lambda$. 
Therefore,  the degeneracy of ground states can be tunned in a much flexible manner. 
The mean values of $\mathbf{F}$ and $\mathbf{L}$ is proportial to $\langle \mathbf{J}\rangle$, and can be computed \cite{rose1995elementary} as 
\bea
\langle \mathbf{F}\rangle = \frac{1-\alpha}{2}\langle \mathbf{J}\rangle, \hspace{.2cm}  \langle \mathbf{L}\rangle = \frac{1+\alpha}{2}\langle \mathbf{J}\rangle, \hspace{.2cm} \langle \mathbf{J}\rangle=j_z\textbf{e}_z,
\eea
with $\alpha= [l(l+1)-f(f+1)]/j(j+1)$.
Therefore, $|\avg(\mathbf{F})|$ and $|\avg(\mathbf{L})|$ can take fractional values for different $j_z$.

The above construction of non-Abelian SOAM coupling in cold atoms provides an avenue to explore various novel physics with high flexibility. First, various spin-momentum coupled subsystem in 2D can be easily obtained by cutting the system appropriately, as shown in \cite{li2012two,li2013high}.
For instance, for fixed $z=z_0$, we have a 2D subsystem with Rashba-type SMC. 
Meanwhile, quantum Hall physics in 3D space can also be induced by non-Abelian SOAM coupling \cite{li2013topological,li2013high}.
Second, the implementation of a spherical surface trap with tunable radius allows us to modify the strength of SOAM coupling, together with the degeneracy of the ground states subspace in a much flexible way.
This also enables the investigation of strong-correlated physics with only a few particles.   
Finally, for spinor condensates, this non-Abelian SOAM coupling results in various spin vortices with intrinsic topology, as will be discussed in the following.

%\section{\label{sec:phase}phase diagram}
\section{Phase diagram}

%The low-energy s-wave contact scatterings for spinor condensates take different forms depending on the spin-$f$.When spin $f=1$, 
For spinor condensates $F=1$ with low-energy s-wave contact scattering,
the field operator $\psi(\theta,\phi)$ contains three components $[\psi_{1}(\theta,\phi)~\psi_{0}(\theta,\phi)~\psi_{-1}(\theta,\phi)]^{T}$, and the reduced contact interaction in the spherical surface is
\begin{align}
	H_{\text{int}}=\int d\Omega~[c_0:\hat{n}^2(\theta,\phi):+c_1:\hat{F}^2(\theta,\phi):].
\end{align}
Here $d\Omega = \sin\theta d\theta d\phi$, and $::$ represents the normal order  of the operator.
$\hat{n}(\theta,\phi)=\sum^{1}_{f_z=-1}\psi^{\dagger}_{f_z}\psi_{f_z}$ is the total particle number. $\hat{F}^2=(\hat{F}_+\hat{F}_-+\hat{F}_-\hat{F}_+)/2+\hat{F}^2_z$ with $\hat{F}_+=\hat{F}^{\dagger}_-=\sqrt{2}(\psi^{\dagger}_1\psi_0+\psi^{\dagger}_0\psi_{-1})$ and $\hat{F}_z=\psi^{\dagger}_1\psi_1-\psi^{\dagger}_{-1}\psi_{-1}$.
$c_0$ and $c_1$ define the reduced dimensionless strengths of density-density and spin-dependent interactions on the surface trap, whose explicit forms can be written as
\bea
	c_0&=&\frac{g_0+2g_2}{6\epsilon_0}N\int r^2\vert\varphi(r)\vert^4~dr\approx \frac{2\sqrt{2\pi}}{3}\frac{a_0+2a_2}{l_T}N,~ \nn\\
	c_1&=&\frac{g_2-g_0}{6\epsilon_0}N\int r^2\vert\varphi(r)\vert^4~dr\approx \frac{2\sqrt{2\pi}}{3}\frac{a_2-a_0}{l_T}N.\nn
\eea
Here $g_0=4\pi\hbar^2a_0/\mu$ and $g_2=4\pi\hbar^2a_2/\mu$ represent the interactions in two-body scattering channels with total spin $F=0$, and $F=2$ respectively. 
$a_0$ and $a_2$ are corresponding s-wave scattering length. 
$N =\int d\Omega \hat{n}(\theta,\phi)$ is the total number of particles.

In the mean-field level, we find the phase diagrams using both the imaginary-time-evolution and variational methods.
Since the single-particle eigen-states are degenerate, the ground states $\psi(\theta,\phi)$ exhibit complex spin patterns even for condensates with weak contact interaction.
When $f=1$, the interaction energy is
\begin{align}
	E_{\text{int}}=\int d\Omega~[c_0  n(\theta,\phi)^2+c_1\vec{\mathcal{F}}(\theta,\phi)^2].
\end{align}
Here $n(\theta,\phi)=|\psi(\theta,\phi)|^2$ is the local density and $\vec{\mathcal{F}}(\theta,\phi)=\psi^{\dag} \mathbf{F} \psi$ represents the local spin-density vector.
Due to the symmetry, the ground state $\psi(\theta,\phi)$ is equivalent up to a global rotation defined as $\mathcal{R}(\alpha,\beta,\gamma)=\exp(-i J_z\alpha)\exp(-i J_y\beta)\exp(-i J_z\gamma)$.

%%%%%%%%%%%%%%%%%%%%%%%%%%%%%%%%%%%%%%%%%
\begin{figure}[ht]
	\includegraphics[width=0.45\textwidth]{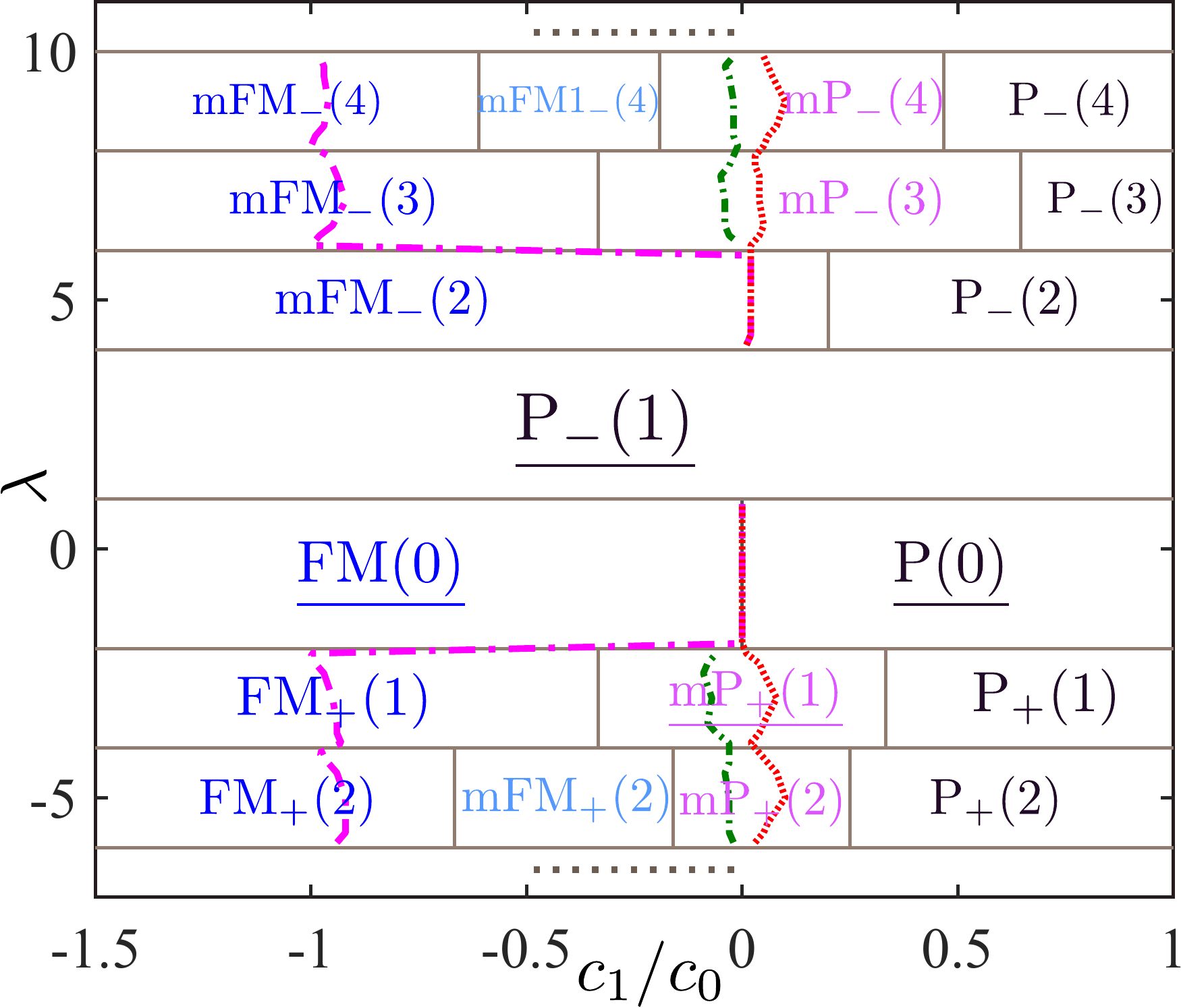}
	% Here is how to import EPS art
	\caption{\label{fig:pdf1}
	The mean-field ground-states of the condensates in the $c_1/c_0-\lambda$ plane. Here the phase boundaries for $c_0=1$ are plotted with solid grey lines. ``FM",``P" and ``m" are short for ferromagnetic, polar and meta respectively. In symbol ``xx($l$)", $l$ is the orbital-angular-momentum (OAM) quantum number of the condensates. ``+" and ``-" indicate that the spin $\textbf{F}$ is parallel or anti-parallel with the OAM $\textbf{L}$. Underlined text means that the phase supports homogeneous density distributions over the surface.  Other dashed lines show the boundaries obtained for large $c_0=100$: the red dotted line represents the boundary between polar states and mP phases; the green dot-dashed lines separate the mP phases and other spin-vortex phases with $|\avg(F)|\neq0$; the boundary for FM phases are depicted with magenta dot-dashed lines.  }
\end{figure}
%%%%%%%%%%%%%%%%%%%%%%%%%%%%%%%%%%%%%%%%%%%%

\subsection{Phase diagram for weak interaction $c_0=1$}

Fig.\ref{fig:pdf1} shows the phase diagram in the $c_1/c_0-\lambda$ plain for small quantum numbers of $l$ and $j$, where the explicit ground-state configurations within different regimes of $\lambda$ are also provided in Tab. 1 and Appendix C.
In these cases, the ground states can be determined quantitatively as the single-particle eigenstates have much lower degeneracy.
The phase diagram shows many novel features, which will be discussed below.

First, due to the presence of interaction, the degeneracy of $\psi^{l,f}_{j,j_z}$ with different $j_z$ is broken even in the presence of very weak interaction $c_0=1$.
For given $\lambda$ and $j$, the ground state carries different $\avg(\mathbf{J})$ (or $\avg(\mathbf{L})$ and $\avg(\mathbf{F})$), and supports various exotic spin patterns depending on the ratio $c_1/c_0$ (See Tab. I for details). 
Meanwhile, the phase diagram exhibits similar structures for the same $j$, regardless of whether $j=l+f$ or $j=|l-f|$.
Therefore, the whole diagram shows an approximate mirror symmetry around the $P_-(1)$ phase with $j=0$ ($l=1$) and $\lambda \in (1,4)$.

\begin{figure}[htbp]
	\includegraphics[width=0.45\textwidth]{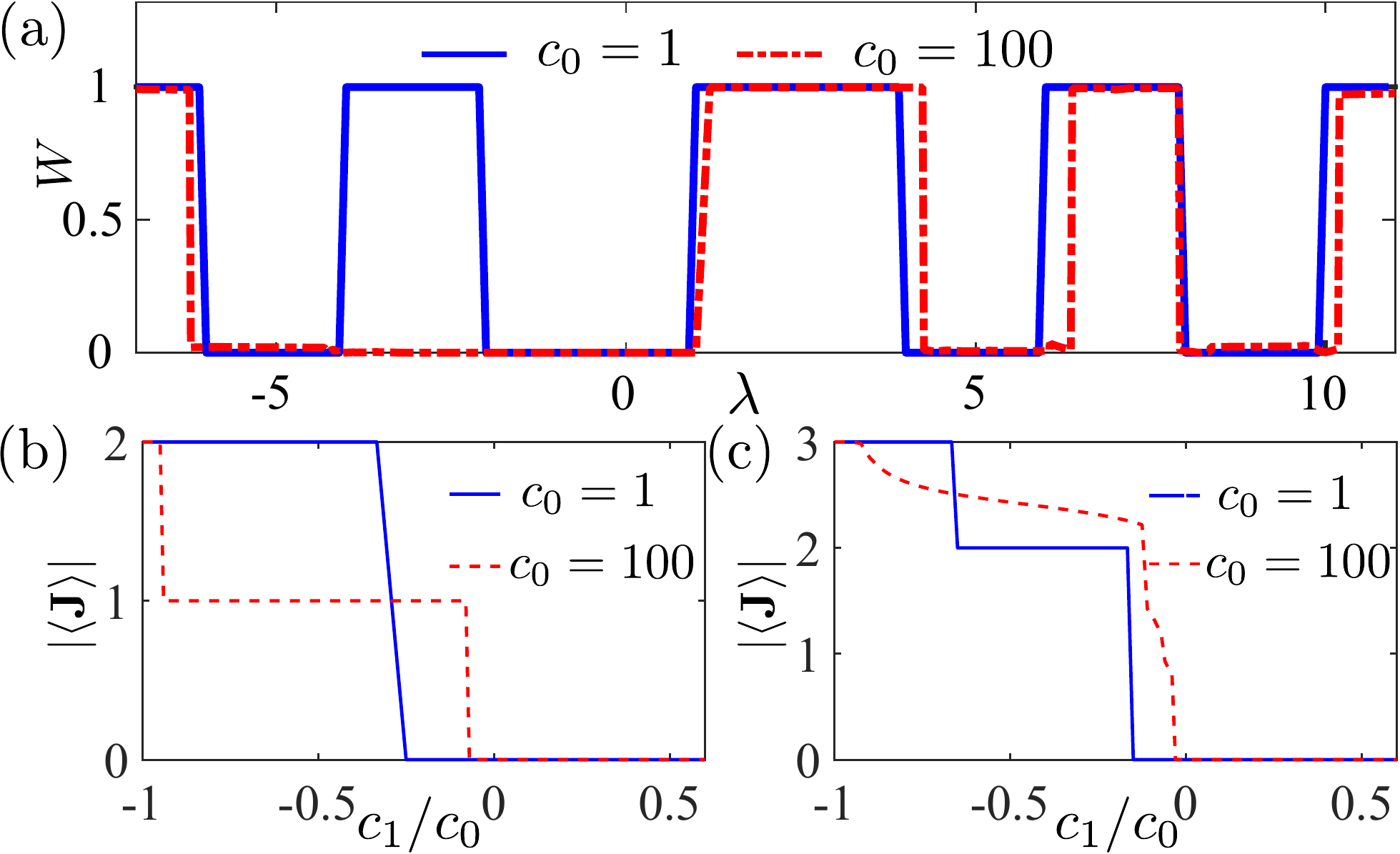}% Here is how to import EPS art
	\caption{\label{fig:meanJ}
	(a) $Z_2$-type topological charge $W$ of polar states along with $\lambda$ for fixed $c_1/c_0=1$, which is closely linked with the parity of the wavefunction. See text for details;	
	(b) and (c) shows the mean values of $\mathbf{J}$ along with ratio $c_1/c_0$ for different $\lambda=-3$ and $\lambda=-5$ respectively. The quantized feature of $|\langle\mathbf{J}\rangle|$ can be stabilized or spoiled for larger $c_0$.	
	} 
\end{figure}

Second, in the abscence of SOAM coupling (see Fig. \ref{fig:pdf1} for regimes with $l=0$ and $\lambda \in (-2,1)$), it is well-known that the condensates can be in the FM and polar phases depending on the sign of $c_1$, where the magnitude of normalized local vector $|\vec{\mathcal{F}}(\theta,\phi)|/n(\theta,\phi)$ takes the value $1$ and $0$ respectively. 
However, the presence of SOAM coupling can result in new ground states with $l \neq 0$, which supports novel vortex patterns and spin textures.  
In addition, the abovementioned FM and polar phases appear only for strong spin-exchanging interaction $|c_1| \sim c_0$. 
These phases also exhibit many new features, which are listed as follows: 
\begin{enumerate}
 	\item $c_1/c_0 \sim -1$ with $\lambda<0$ and $j=l+f$. 
 	In this case, the ground-state wavefunction reads $\psi^{l,f}_{j,j}(\theta,\phi)$. Sicne $j_z=j$, this corresponds to the usual FM phases with maximized local vector $\vec{\mathcal{F}}(\theta,\phi)$ satisfying $|\vec{\mathcal{F}}(\theta,\phi)|/n(\theta,\phi)=1$.  
In addition, the spin fluctuation defined by $\Delta\mathcal{F}_{ij}= \langle \mathcal{F}_{i}\mathcal{F}_j \rangle -\langle \mathcal{F}_{i}\rangle\langle \mathcal{F}_{j}\rangle$ with $(i,j) \in (x,y,z)$ also vanishes. These states are also denoted by FM$_+(l)$ 
 	\item  $c_1/c_0 \sim -1$ with $\lambda>0$ and $j=|l-f|$. Here the ground states also reads $\psi^{l,f}_{j,j_z=j}(\theta,\phi)$. Since $j<l$, the normalized vector $|\vec{\mathcal{F}}(\theta,\phi)|/n(\theta,\phi)<1$ and changes its direction over the surface. Therefore, the state possesses nonzero the spin fluctuation $\Delta\mathcal{F} \neq 0$, and is then called as the mFM phase.
	\item  $c_1/c_0 \sim 1$ for all $\lambda$. In this case, the system supports various polar states with $\vec{\mathcal{F}}(\theta,\phi)=0$, which exhibits novel intrinsic topological properties. To show this, we write the wavefunction using the real nematic order $\vec{d}$ as $
\psi(\theta,\phi) = \vec{d}\cdot |\vec{r}\rangle = d_x|x\rangle + d_y|y\rangle + d_z|z\rangle$,
where $\{|x\rangle, |y\rangle, |z\rangle \}$ are the Cartesian basis formed by the eigenvectors of $F_{x,y,z}$ with zero eigenvalues \citep{kawaguchi2012spinor,zibold2016spin}. 
For a close spherical surface, the unit vector $\hat{d}=\vec{d}/|\vec{d}|$ exhibits nontrivial distribution which can be described by the topological charge $W$ as 
\bea
 %W = \left|\int d\Omega \hat{d}\cdot (\hat{d}_{\theta} \times \hat{d}_{\phi}) \right|,
 W =\frac{1}{4\pi} \left|\int d\theta~ d\phi~ \hat{d}\cdot (\partial_{\theta}\hat{d} \times \partial_{\phi}\hat{d}) \right|,
\eea
where we have introduced the absolute value to avoid the global ambiguity of $\vec{d}$ and $-\vec{d}$.
Fig. \ref{fig:meanJ}a shows that only two values $0$ and $1$ are allowed for the charge $W$. 
This $Z_2$ feature of $W$ is directly related to the parity of $\vec{d}(\theta,\phi)$, as we have $\vec{d}(\pi-\theta,\pi+\phi) \rightarrow (-1)^l \vec{d}(\theta,\phi)$ due to $Y^m_l(\pi-\theta,\pi+\phi) \rightarrow (-1)^l Y^m_l(\theta,\phi)$.
		Especially, when $j=0$, the polar phase $\text{P}_-(1)$ survives for all $c_1/c_0$, and the relevant vector $\hat{d}$ exhibits a stable hedge-hog like pattern with nonzero topological charge $W=1$.
		We note that for condensates without SOAM coupling, such polar state is unstable towards the formaiton of Alice ring, as shown in \cite{kawaguchi2012spinor}.
\end{enumerate} 
  
%%%%%%%%%%%%%%%%%%%%%%%%%%%%%%%%%%%%%%%%%%%%%%%%%%%%%%%%%%%%%%%%%%%%%%%%%
%%%%%%%%%%%%%%%%%%%%%%%%%%%%%%%%%%%%%%%%%%%%%%%%%%%%%%%%%%%%%%%%%%%%%%%%%  

%%%%%%%%%%%%%%%%%%%%%%%%%%%%%%%%%%%%%%%%%%%%%%%%%%%%%%%%%%%%%%
\begin{table*}
\renewcommand\arraystretch{1.5}
\centering
\caption{Explicit information of different phases in figure 2 for $\lambda <0$ within different regimes $c_1/c_0$. Here "WF" is short for "wavefunction". Other symbols are defined as follows: $\eta=(41-\sqrt{5233})/192 \simeq -0.16 $, $\psi_a=[\sqrt{2}\psi^{1,1}_{2,-1}+\psi^{1,1}_{2,2}]/\sqrt{3}$, $\psi_c=[\psi^{2,1}_{3,2}+\psi^{2,1}_{3,-2}]/\sqrt{2}$, and $\psi_b=\alpha \left [ \psi^{2,1}_{3,-3}+\psi^{2,1}_{3,3} \right ] +\beta \psi^{2,1}_{3,0}$  with $\alpha=\sqrt{(11c_0-20c_1)/(47c_0-80c_1)}$ and $\beta=\sqrt{1-2\alpha^2}$.}
\begin{tabular}{||c||c|c||c|c|c||c|c|c|c||}
  \hline
    & \underline{FM(0)} & \underline{P(0)} & FM$_+(1)$ & \underline{mP$_+(1)$} & P$_+(1)$ & FM$_+(2)$ & mFM$_+(2)$ & mP$_+(2)$ & P$_+(2)$ \\
  \hline
   $c_1/c_0$   & $(-\infty,0)$ & $(0,\infty)$ & $(-\infty,-1/3)$ & $(-1/3,1/3)$ & $(1/3,\infty)$ & $(-\infty,-2/3)$ & $(-2/3,\eta)$ & $(\eta,1/4)$ & $(1/4,\infty)$ \\
  \hline
  WF & $\psi^{0,1}_{1, 1}$ & $\psi^{0,1}_{1,0}$ & $\psi^{1,1}_{2,2}$ & $\psi_a$ & $\psi^{1,1}_{2,0}$ & $\psi^{2,1}_{3,3}$ & $\psi^{2,1}_{3,2}$ & $\psi_b$ & $\psi_c$ \\
  \hline
  $(|\avg(\mathbf{J})|,|\avg(\mathbf{F})|)$ & $(1,1)$ & $(1,0)$ & $(2,1)$ & $(0,0)$ & $(0,0)$ & $(3,1)$ & $(2,2/3)$ & $(0,0)$ & $(0,0)$ \\
  \hline  
  $\Delta \mathcal{F}$ & $0_{3\times3}$ & $0_{3\times3}$ & $0_{3\times3}$ & $\frac{\text{diag}\{1,1,1\}}{20\pi}$ & $0_{3\times3}$ & $0_{3\times3}$ & $\frac{\text{diag}\{5,5,20\}}{126\pi}$  & $\frac{\text{diag}\{3,3,2\}\beta^2\alpha^2}{7\pi}$ & $0_{3\times3}$ \\
%  \hline
%  $E_{int}$ & x  &  x & x  & $\frac{5c_0+3c_1}{20\pi}$ & x & x & $\frac{5(7c_0+6c_1)}{126\pi}$ & x & $\frac{25c_0}{84\pi}$ \\
  \hline
\end{tabular}
\label{WF:information}
\end{table*}
%%%%%%%%%%%%%%%%%%%%%%%%%%%%%%%%%%%%%%%%%%%%%%%%%%%%%%%%%%%%%%%%%%%%%%%%%

Finally, across the intermediate regimes of $c_1/c_0 \in (-1, 1)$, the system transits between the FM and polar states, and various new phases arise.
These phases support quantized mean values of $|\avg(\mathbf{J})|$ (or, $|\avg(\mathbf{L})|$, $|\avg(\mathbf{F})|$), as shown in Fig. \ref{fig:meanJ}(b-c), which represents another key feature of such SOAM coupled condensates.
Since both $|\avg(\mathbf{F})|$ and the fluctuation $\Delta\mathcal{F}$ take nonzero values, they still belong to the mFM phases.
Beside this, the system also supports another mP states with $|\avg(\mathbf{J})|=|\avg(\mathbf{F})|=0$ and nonzero local spin-density vector $\vec{\mathcal{F}}(\theta,\phi)\neq 0$.
For instance, when $j=2$, $\lambda\in(-4,-2)$, and $c_1/c_0 \in (-1/3,1/3)$,  the mP state reads
\[\psi_a(\theta,\phi)=[\sqrt{2}\psi^{1,1}_{2,-1}(\theta,\phi)+\psi^{1,1}_{2,2}(\theta,\phi)]/\sqrt{{3}} \] 
and supports a homogeneous density distribution over the spherical surface with isotropic spin fluctuation $\Delta\mathcal{F} \sim I_{3\times3}$.
While for $c_1/c_0>1/3$, polar phase with non-homogeneous distribution is preferred so that the spin-dependent interaction is minimized.
We note that all the abovementioned transitions are of first-order.

%%%%%%%%%%%%%%%%%%%%%%%%%%%%%%%%%%%%%%%%%%%%%%%%%%%%%%%%%%%%%%%%%%%%%%%%%
%\begin{figure}[htbp]
%	\includegraphics[width=0.35\textwidth]{./figure/chernnb.eps}% Here is how to import EPS art
%	\caption{\label{fig:chernnb}
%	$Z_2$-type topological charge $W$ of polar states along with $\lambda$ for fixed $c_1/c_0=1$, which is closely linked with the parity of the wavefunction. See text for details.	
%	}
%\end{figure}
%%%%%%%%%%%%%%%%%%%%%%%%%%%%%%%%%%%%%%%%%%%%%%%%%%%%%%%%%%%%%%%%%%%%%%%%%

\subsection{Phase diagram for strong interaction $c_0=10^2$}

For larger contact interaction $c_0=100$, eigenstates $\psi^{l,f}_{j,j_z}$ with different $j$ can be mixed to form new ground states so that the density distribution of the condensates becomes more uniform.
This leads to quantitative changes of all the previous results.

First, the boundaries for the FM and polar phases move leftwards on the $c_1/c_0-\lambda$ plane for all phases with $l \neq 0$, as shown in Fig. \ref{fig:pdf1} with magenta dot-dashed and red dotted lines respectively. 

Second, topological charge $W$ defined in the polar regimes around $c_1/c_0=1$ shift and even vanishes when $\lambda \in (-4,-2)$, as shown in Fig.\ref{fig:meanJ}a. 
This is a direct evidence that the ground state can no longer be written as the superposition of different $\psi^{l,f}_{j,j_z}$ with fixed $j=2$ and $l=1$.  Additional components with different $j$ and $l$ should be involved.

Finally, the regime of mP phases shrinks for large $c_0$ and distribute mainly around the line with $c_1=0$.
On the other hand, the mFM regimes in the phase diagram becomes larger.
These intermediate mFM phases show complex patterns, and can not be simply characterised using quantized $\avg(\mathbf{J})$ (or $\avg(\mathbf{L})$ and $\avg(\mathbf{F})$) any more.
For instance, new mFM phase with fixed $|\avg(\mathbf{J})|=1$ appears for intermediate $c_1/c_0$, as shown in Fig. \ref{fig:meanJ}b. 
While for larger $|\lambda|=5$, the quantized feature of  $|\avg(\mathbf{J})|$ breaks (Fig.\ref{fig:meanJ}c), which makes the discrimination of different mFM states to be a numerically challenging task. We leave this for further investigation.

\subsection{Thomson lattices formed by topological defects}

%%%%%%%%%%%%%%%%%%%%%%%%%%%%%%%%%%%%%%%%%%%%%%%
\begin{figure}[htbp]
	\includegraphics[width=0.48\textwidth]{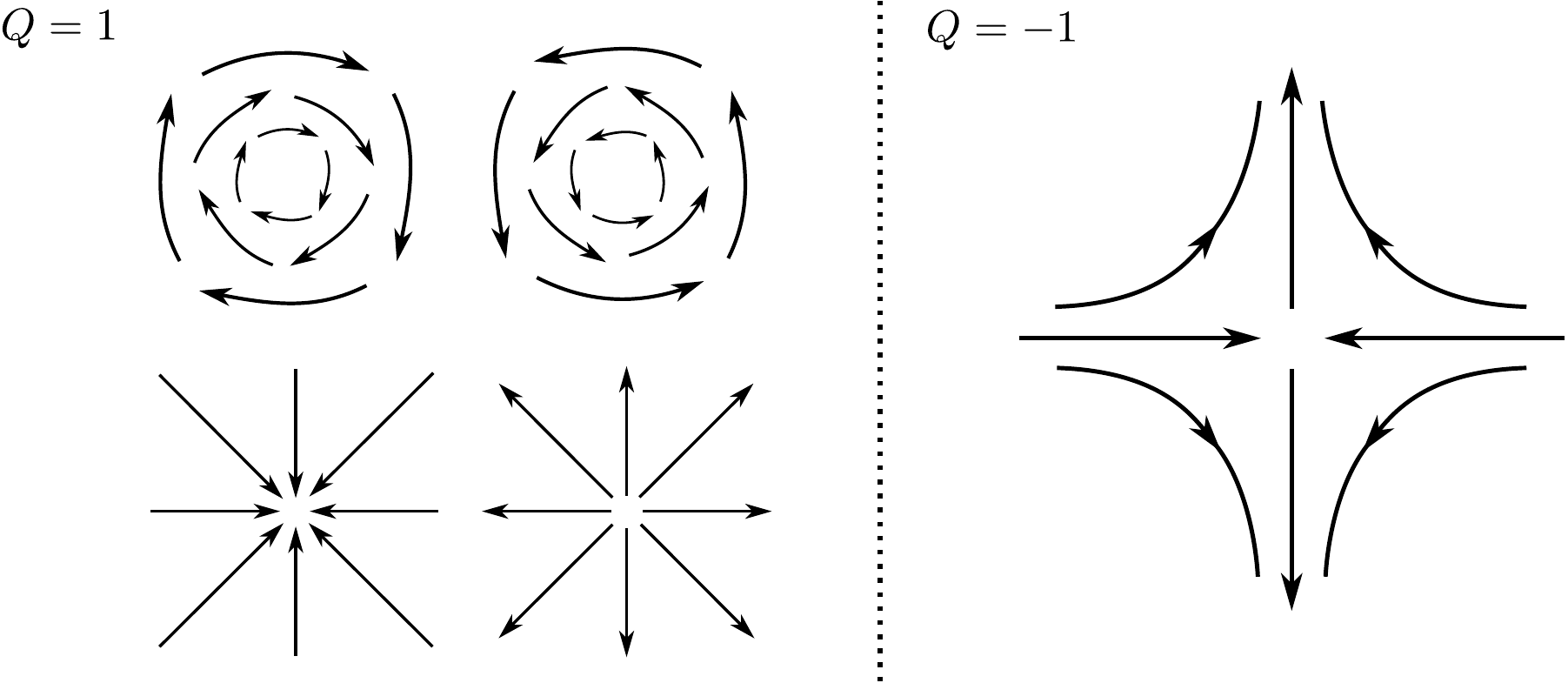}% Here is how to import EPS art
	\caption{%\label{fig:sphe_trap1}
		Spin vortex defects on the spherical surface in the meta-polar phase, where only defects with two possible Poincar\'{e} index $Q= +1$ and $-1$ are numerically obtained.} \label{fig:topQ0}
\end{figure}
%%%%%%%%%%%%%%%%%%%%%%%%%%%%%%%%%%%%%%%%%%%%%%%%%%%%

In the case of neglectable spin-exchanging interactions around $c_1/c_0 \sim 0$, which is fulfilled in most current experiments, the condensates spread almost homogeneously over the surface so that the contact interaction is minimized. 
The local vector $\vec{\mathcal{F}}(\theta,\phi)$ changes its magnitude and direction around the closed surface, with its tangential component forming different-types of defects. 
These defects can be characterised using the topological Poincar\'{e} index $Q$, which only takes the value $+1$ or $-1$ in our case, as shown in Fig. \ref{fig:topQ0}.
The number of topological defects $N_{\pm}$ with different $Q=\pm1$ satisfies the Poincar\'{e}-Hopf theorem 
\[N_+-N_- = 2,\]
which comes from the boundaryless feature of such surface trap. 
Around each defect center, the spin texture forms a coreless vortex \citep{kawaguchi2012spinor,mizushima2004coreless,lovegrove2016stability}.
%% \red{Fig.xx} shows the typical spin texture formed on the surface for different $\lambda$.
Interestingly, for $Q=-1$, we always have polar-core spin vortices with $\vec{\mathcal{F}}(\theta,\phi)=0$ at the center. 
While for $Q=+1$, we have coreless FM-centered vortices with the nomalized vector $|\vec{\mathcal{F}}(\theta,\phi)|/n(\theta,\phi) \sim 1$, or coreless mFM-centered vortices with $|\vec{\mathcal{F}}|/n < 1$ (See Appendix D for details).  

%%%%%%%%%%%%%%%%%%%%%%%%%%%%%%%%%%%%%%%%%%%%%%%%%%%%%%%%%%%%%%%%%%%%%%%%%
\begin{figure}[htbp]
	\includegraphics[width=0.48\textwidth]{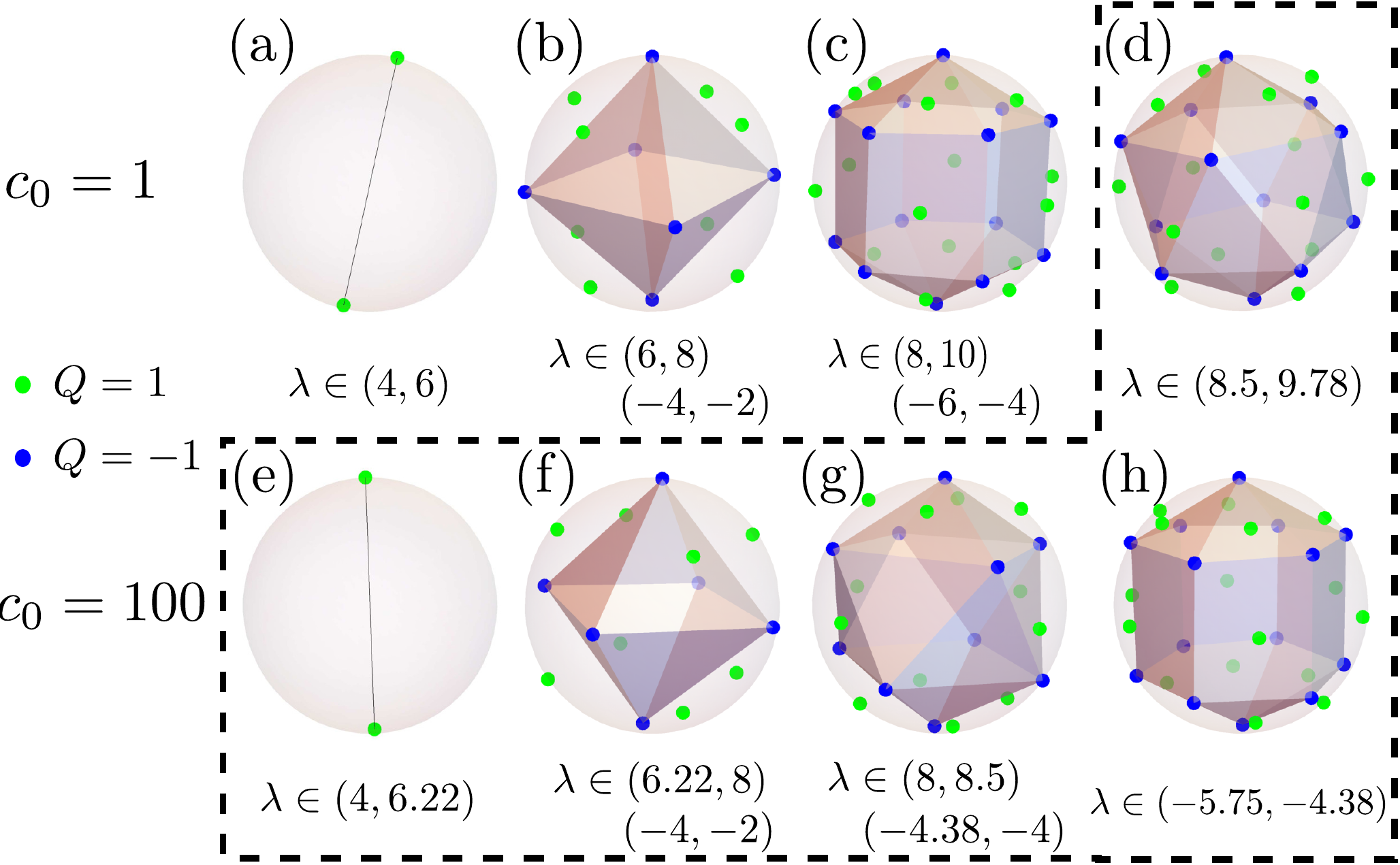}% Here is how to import EPS art
	\caption{\label{fig:spinvortex}
	Topological defects of spin vortices on spherical surface with charge $Q=+1$ and $Q=-1$ respectively for different contact interaction $c_0=1$ and $100$ respectively. In all cases, defects with $Q=-1$ form the standard Thomson lattices. 
	}
\end{figure}
%%%%%%%%%%%%%%%%%%%%%%%%%%%%%%%%%%%%%%%%%%%%%%%%%%%%%%%%%%%%%%%%%%%%%%%%%

To minimize the interaction effect, these vortex defects form regular patterns around the spherical surface, as depicted in Fig. \ref{fig:spinvortex}. 
In all cases, the defect with $Q=-1$ and $\vec{\mathcal{F}}(\theta,\phi)=0$ form stable configurations characterised by the solution of Thomson problem  for $N_-$ electrons \citep{thomson1904xxiv,zhou2018synthetic}.  
This also verifies the well-known charge-vortex duality for magnetic vortices in 2D system.
In the case of small $c_0=1$, the Thomson lattices are the same when the single-particle eigenstates share the same TAM $j$ for given $\lambda$, as shown in Fig.\ref{fig:spinvortex}(a-c).
However, for larger $c_0=100$, the ground states can be the superposition of eigenstates with different $j$ . This results in new lattice patterns for different $\lambda$, as depicted in figure \ref{fig:spinvortex}(d-h). 
We note that except the special case with $N_+=2$ and $N_-=0$, no such elegant Thomson pattern has been found for defects $Q=+1$ with nozero local vector $\vec{\mathcal{F}}(\theta,\phi)$.

%The ground states of the condensates read $\psi^{0,1}_{j,j_z}(\theta,\phi)=Y_{00}(\theta,\phi)\otimes\chi$ and share similar properties as those without SOAM.
%When $c_1/c_0<0$, we obtain a ferromagnetic state with $\chi=(1,0,0)^T$, in which all spins point along the $z$-axis with the mean value $\vert\mathbf{F}\vert\equiv\vert\langle\mathbf{F}\rangle\vert=1$.
%On the contrary, we have the polar phase with $\chi=(0,1,0)^T$ and $\vert\mathbf{F}\vert=0$.

%\section{\label{sec:con}conclusion}
\section{Experimental consideration and Conclusion}
For $^{87}$Rb atoms which has been widely studied in current experiments, the relevant parameters chosen in the paper are summaried as follows: specifically, by choosing suitable $U_0$, $U$ and $w_1$ we can have a spherical surface trap with the spherical radius $R=10~\mu$m $\gg l_T\approx 0.48~\mu$m.
The oscillating frequency of magnetic gradient $B_1$ can be set to satisfy $\Omega_0=2\pi\times 50$ kHz $\gg \omega=2\pi\times 0.5$ kHz. In this case,
the strength of SOAM coupling reads $\lambda\in(0.98,15.7)$ when $B_1\in(50,200)$ G/cm.
The constant bais reads $B_0=6.0$ G, which ensures that the following relation $\omega_0=2\pi\times 4.2$ MHz $\gg \Omega_0$ is met.
In concrete experiments, the density $\rho$ of $^{87}\text{Rb}$ BECs can be $10^{11}$ cm$^{-3}$ to $10^{13}$ cm$^{-3}$.
The total number of particles $N\sim 4\pi R^2 2l_T\rho$ in our spherical trap could reach to $10^2$ to $10^{4}$. 
The dimensionless interactions $c_0$ can reach $10^0-10^2$, as required by our calculations.

To summarize, we have proposed an promising route to explore non-Ableian SOAM coupling in cold atomic system with the help of synthetic monopole fields.
The flexibility of the system allows us to construct an effective spherical surface trap, where its ground-state degeneracy can be tuned in a wide parameter regimes.
For spinor condensates with $f=1$, we show that the system supports various exotic mFM, mP, and polar phases with nontrivial intrinsic topology.
The proposed method works for both bosons and fermions, which thus opens up an avenue to explore various spin vortices on curved surfaces, and may provide a new routine to investigate strong-correlated physics using ultra-cold atoms with tunable ground-state degeneracy.

\begin{acknowledgments}
	XFZ thanks Congjun Wu, Yi Li, and Shao-Liang Zhang for many helpful discussions.
	This work was funded by National Natural Science Foundation of China (Grants No. 11774332, No. 11774328, No. 11574294, and No. 11474266), the major research plan of the NSFC (Grant No. 91536219),the USTC start-up funding (Grants No. KY2030000066, No. KY2030000053), the National Plan	on Key Basic Research and Development (Grant No. 2016YFA0301700), and the Strategic Priority Research Program (B) of the Chinese Academy of Sciences (Grant No. XDB01030200).
	M.G. also thanks the support by the National
Youth Thousand Talents Program (No. KJ2030000001).
%	J.M.C acknowledges computer-hardware support from Ming Gong.
\end{acknowledgments}

\appendix

\vspace{1cm}

\section{Spin-orbital-angular-momentum coupling with negative sign} \label{app:soam_ns}
The realization of the SOAM coupling with negative coefficient using gradient magnetic pulses can be divided into two steps.

%%%%%%%%%%%%%%%%%%%%%%%%%%%%%%%%%%%%%%%%%%%%%%%%%%%%%%
\begin{figure}[htbp]
	\includegraphics[width=0.45\textwidth]{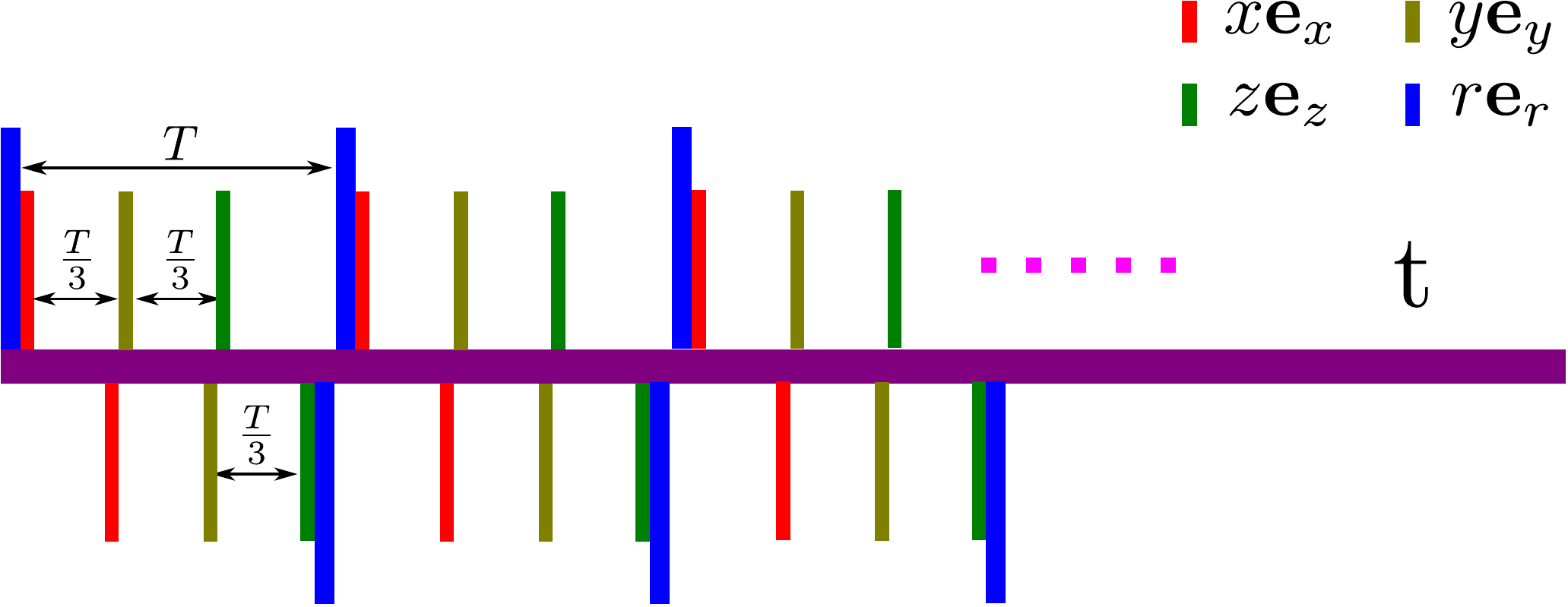}% Here is how to import EPS art
	\caption{\label{fig:mod}
	Magnetic field modulation scheme of negative SOAM coupling.}
\end{figure}
%%%%%%%%%%%%%%%%%%%%%%%%%%%%%%%%%%%%%%%%%%%%%%%%%%%%

First, using the standard magnetic pulses, such as $\mathbf{B}\propto x \mathbf{e}_x, y\mathbf{e}_y, z\mathbf{e}_z$, we can implement an intermediate 3D spin-momentum coupling as
\bea
	H'_{\text{eff}} \simeq H_0-\frac{\lambda^{\prime}\tau}{3\mu}\mathbf{F}\cdot\mathbf{P}.
\eea
\begin{widetext}
This is possible if we consider the following sequence of magnetic pulses as
\begin{flalign}
\mathcal{U'}(T)=\exp[-i\frac{T}{\hbar}H'_{\text{eff}}]=&\left(e^{i\frac{\lambda^{\prime}\tau}{\hbar}zF_z} e^{-i\frac{T}{3\hbar}H_0} e^{-i\frac{\lambda^{\prime}\tau}{\hbar}zF_z} \right )
	\left(e^{i\frac{\lambda^{\prime}\tau}{\hbar}yF_y} e^{-i\frac{T}{3\hbar}H_0} e^{-i\frac{\lambda^{\prime}\tau}{\hbar}yF_y} \right )
	\left(e^{i\frac{\lambda^{\prime}\tau}{\hbar}xF_x} e^{-i\frac{T}{3\hbar}H_0} e^{-i\frac{\lambda^{\prime}\tau}{\hbar}xF_x}  \right ) \notag \\
%	=&\exp[-i\frac{T}{3\hbar}e^{i\frac{\lambda^{\prime}\tau}{\hbar}zF_z}H_0e^{-i\frac{\lambda^{\prime}\tau}{\hbar}zF_z}] 	\exp[-i\frac{T}{3\hbar}e^{i\frac{\lambda^{\prime}\tau}{\hbar}yF_y}H_0e^{-i\frac{\lambda^{\prime}\tau}{\hbar}yF_y}]	\exp[-i\frac{T}{3\hbar}e^{i\frac{\lambda^{\prime}\tau}{\hbar}xF_x}H_0e^{-i\frac{\lambda^{\prime}\tau}{\hbar}xF_x}] \notag \\
	\approx&\exp[-i\frac{T}{3\hbar}(H_0-\frac{\lambda^{\prime}\tau}{\mu}F_zP_z)]
	 \exp[-i\frac{T}{3\hbar}(H_0-\frac{\lambda^{\prime}\tau}{\mu}F_yP_y)]
	 \exp[-i\frac{T}{3\hbar}(H_0-\frac{\lambda^{\prime}\tau}{\mu}F_xP_x)]  \notag \\
	\approx&\exp[-i\frac{T}{\hbar}(H_0-\frac{\lambda^{\prime}\tau}{3\mu}\mathbf{F}\cdot\mathbf{P})],
\end{flalign}
where we have assumed that the magnetic pulse is strong enough so that during the time interval $\tau \ll T$, the free evolution of the system can be neglected.

The second step is employing the hedge-hog like magnetic pulses to realize the desired SOAM coupling.
The corresponding evolution operator reads
\begin{flalign}
\mathcal{U}(T)=&\exp[-i\frac{T}{\hbar}H_{\text{eff}}]= e^{i\frac{\lambda\tau}{\hbar}\mathbf{r}\cdot\mathbf{F}} \mathcal{U'}(T) e^{-i\frac{\lambda\tau}{\hbar}\mathbf{r}\cdot\mathbf{F}} \approx e^{i\frac{\lambda\tau}{\hbar}\mathbf{r}\cdot\mathbf{F}}\exp[-i\frac{T}{\hbar}(H_0-\frac{\lambda^{\prime}\tau}{3\mu}\mathbf{F}\cdot\mathbf{P})]e^{-i\frac{\lambda\tau}{\hbar}\mathbf{r}\cdot\mathbf{F}} \notag \\
\approx&\exp\left[-i\frac{T}{\hbar}(H_0-\frac{(\lambda^{\prime}+3\lambda)\tau}{3\mu}\mathbf{F}\cdot\mathbf{P}+\frac{\tau^2\lambda}{\mu} (\frac{\lambda^{\prime}}{3}+\frac{\lambda}{2})(\mathbf{L}\cdot\mathbf{F}+\mathbf{F}^2))\right].
\end{flalign}
\end{widetext}
So if we set $\lambda^{\prime}=-3\lambda$, then the evolution operator becomes
\begin{align}
	\mathcal{U}(T)= e^{-i\frac{T}{\hbar}H_{\text{eff}}}	\approx\exp \left[-i\frac{T}{\hbar}(H_0-\frac{\tau^2\lambda^2}{2\mu}(\mathbf{L}\cdot\mathbf{F}+\mathbf{F}^2))\right],  \nn
\end{align}
thus, we have that
\begin{align}
	H_{\text{eff}}\approx H_0-\frac{\tau^2\lambda^2}{2\mu}(\mathbf{L}\cdot\mathbf{F}+\mathbf{F}^2),
\end{align}
which,up to a constant term $\mathbf{F}^2$, is the desired SOAM coupled Hamiltonian with negative coupling coefficient.

\section{The reduced model and eigenstates of free Hamiltonian }

We consider a spin-$f$ bosonic gas confined in a spherical surface trap $V(r)$ around $(R-\delta,R+\delta)$ with $\delta \ll R$.
The condensates suffer from a SOAM coupling defined by $\tilde{\lambda}\hbar^2\mathbf{L}\cdot\mathbf{F}$.
For low-energy physics considered here, the radial motion of bosons is frozen and its field operator can be assumed to be $\varphi(r)\psi(\theta,\phi)$. The total Hamiltonian can be divided into two parties
\begin{align}
	\label{eq:htot}
	H_{\text{tot}}=H_0+H_{\text{int}},
\end{align}
in which single-particle Hamiltonian $H_0$ has following form
\begin{equation}
	H_0=\int^{R+\delta}_{R-\delta}r^2d r\int d\Omega \varphi^{\ast}(r)\psi^{\dagger}(\theta,\phi)\tilde{\mathcal{H}}_0\varphi(r)\psi(\theta,\phi),
\end{equation}
where $d\Omega=\sin\theta d\theta d\phi$, and the Hamiltonian $\tilde{\mathcal{H}}_0$ contains the SOAM coupling and reads
\begin{align}
	\tilde{\mathcal{H}}_0=-\frac{\hbar^2}{2\mu r^2}\frac{\partial}{\partial r}(r^2\frac{\partial}{\partial r})+\frac{\hbar^2 \mathbf{L}^2}{2\mu r^2}+\tilde{\lambda}\hbar^2\mathbf{L}\cdot\mathbf{F} +V(r).
\end{align}
Here $\mu$ is atomic mass, $\tilde{\lambda}$ stands for SOAM coupling strength.
After integrating out the radial degree-of-freedom, we obtain a reduced single-particle Hamiltonian in a spherical surface
\begin{align}
	H_0=\frac{\mathcal{B}\hbar^2N}{2\mu R^2}\int d\Omega\psi^{\dagger}(\theta,\phi)\mathcal{H}_0\psi(\theta,\phi)+NE_0,
\end{align}
where $E_0=\int^{R+\delta}_{R-\delta}r^2\varphi^{\ast}(r)[(-\frac{\hbar^2}{2\mu})\frac{\partial}{\partial r}(r^2\frac{\partial}{\partial r})+V(r)]\varphi(r)dr$ is the energy arising from radial motion,  $N$ is the total particle number, and $\mathcal{B}=\int^{R+\delta}_{R-\delta}r^2\vert\varphi(r)\vert^2dr \simeq 1$ due to the normalization of $\varphi(r)$.
We have adopted normalization condition $\int d\Omega\psi^{\dagger}\psi=1$.
Hereafter, we neglect the last term $NE_0$ by selecting a new zero energy point and set $\epsilon_0=\frac{\mathcal{B}\hbar^2}{2\mu R^2}$ as energy unit. The free particle Hamiltonian in the spherical surface with radius $R$ can then read
\begin{align}
	\label{eq:H0}
	\mathcal{H}_0=\mathbf{L}^2+\lambda\mathbf{L}\cdot\mathbf{F},\mbox{ with }
	\lambda=2\mu R^2\tilde{\lambda}.
\end{align}

\begin{figure}[htbp]
	\includegraphics[width=0.42\textwidth]{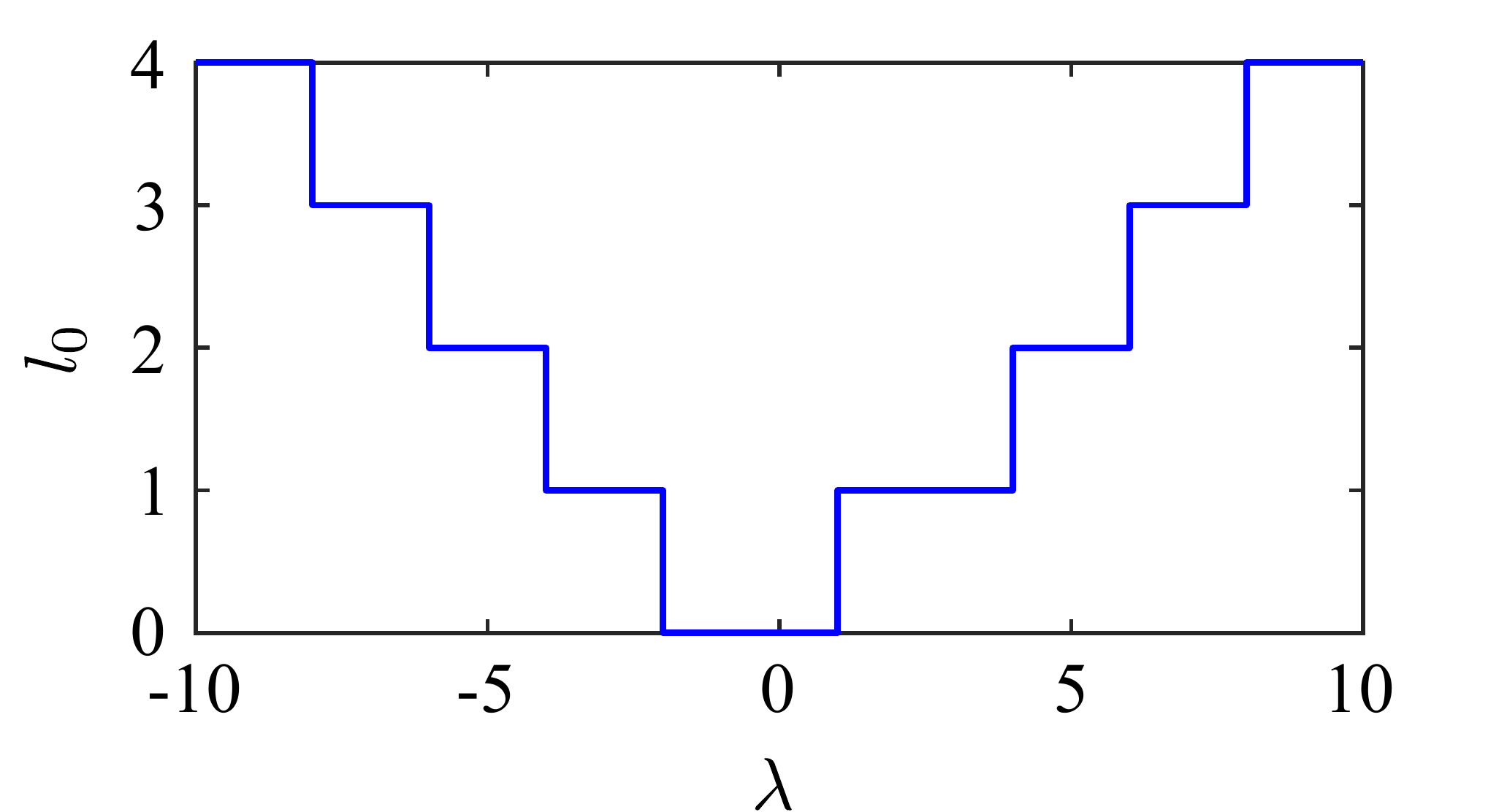}% Here is how to import EPS art
	\caption{\label{fig:angularmomentum_gd}
	Orbital-angular-momentum quantum number $l_0$ of single-particle ground-states verses spin-orbital-angular-momentum coupling $\lambda$ for spin $f=1$.}
\end{figure}

Since the total angular-momentum $\mathbf{J}=\mathbf{L}+\mathbf{F}$ of the system is invariant,  and we have $[\mathcal{H}_0,\mathbf{J}^2]=0$, $[\mathcal{H}_0,\mathbf{L}^2]=0$, and $[\mathcal{H}_0,\mathbf{F}^2]=0$.
According to equality $\mathbf{L}\cdot\mathbf{F}=\frac{1}{2}(\mathbf{J}^2-\mathbf{L}^2-\mathbf{F}^2)$, we can derive its single-particle energy
\begin{align}
	\mathcal{E}_0=l(l+1)+\frac{\lambda}{2}[j(j+1)-l(l+1)-f(f+1)]
\end{align}
with $j=\vert l-f\vert,\vert l-f\vert+1,\cdots, l+f$.
So its ground state configuration is determined by $\lambda$ with the degeneracy given by $2j+1$.
The energy of the lowest energy band for $\lambda > 0$ then reads
\begin{widetext}
\begin{align}
	\mathcal{E}_0=\begin{cases}
		[l-\frac{\lambda(f+1)-1}{2}]^2-\frac{[\lambda(f+1)-1]^2}{4}
		    \text{ \hspace{2cm} if $l\le f$ and $\lambda\in(\frac{2l}{f+1},\frac{2(l+1)}{f+1})$}, \\
		[l-\frac{\lambda f-1}{2}]^2-\frac{(1-\lambda f)^2}{4} -\lambda f
		    \text{ \hspace{2.25cm} if $l\ge f$ and $\lambda\in(\frac{2l}{f},\frac{2(l+1)}{f})$. }
	\end{cases}
\end{align}
When $\lambda<0$, $\mathbf{L}$ is parallel to $\mathbf{F}$, and we have $j=l+f$. The ground states energy is
\begin{align}
	\mathcal{E}_0=(l+\frac{f\lambda+1}{2})^2-\frac{(f\lambda+1)^2}{4} \mbox{ when } \lambda\in(-\frac{2(l+1)}{f},-\frac{2l}{f}).
\end{align}
In this way, we can figure out relations between orbital-angular-momentum quantum number $l_0$ of ground states and the corresponding spin-orbit coupling strength $\lambda$ for spin $f=1$ (see Fig.\ref{fig:angularmomentum_gd}).

\end{widetext}

%%%%%%%%%%%%%%%%%%%%%%%%%%%%%%%%%%%%
%%%%%%%%%%%%%%%%%%%%%%%%%%%%%%%%%%%%%%

Specifically, for spinor condensates with $f=1$, when $\lambda\in \left( -2(l+1),-2l \right) (l\ge1)$, the single-particle ground-states are organized such that $\mathbf{L}$ is parallel with $\mathbf{F}$.
Therefore we have $j=l+1$ with $j_z=-l-1,-l,\cdots,l+1$.
The ground-state is of $2j+1$-fold degeneracy and can be written as
\begin{align}
	\label{eq:psijz2}
	\psi^{l,1}_{l+1,j_z}(\theta,\phi)=\begin{pmatrix}\sqrt{\frac{(l+j_z)(l+j_z+1)}{(2l+1)(2l+2)}}Y_{l,j_z-1}(\theta,\phi) \\
		\sqrt{\frac{2(l-j_z+1)(l+j_z+1)}{(2l+1)(2l+2)}}Y_{l,j_z}(\theta,\phi) \\
	\sqrt{\frac{(l-j_z)(l-j_z+1)}{(2l+1)(2l+2)}}Y_{l,j_z+1}(\theta,\phi) \end{pmatrix}.
\end{align}

When $\lambda\in(-2,1)$, we have $j=f=1$ and $l=0$.  The ground-state is  $2j+1=3$-fold degenerate and reads
\begin{align}
	\psi^{0,1}_{1,j_z}(\theta,\phi)=Y_{0,0}(\theta,\phi)\vert f_z=j_z\rangle.
\end{align}

When $\lambda\in(1,4)$, the OAM $\mathbf{L}$ of the atoms is anti-parallel with $\mathbf{F}$ for the ground states, so we have $j=l-1=0$. The ground-state has no degeneracy and is in form of
\begin{align}
	\label{eq:psi00}
	\psi^{1,1}_{0,0}(\theta,\phi)=\frac{1}{\sqrt{3}}\begin{pmatrix}Y_{1,-1}(\theta,\phi) \\
		-Y_{1,0}(\theta,\phi) \\
	Y_{1,1}(\theta,\phi) \end{pmatrix}.
\end{align}
We also address that this state supports a homogeneous density distribution, and the nematic vector $\vec{d}$ exhibits hedge-hog like pattern over the spherical surface.

When $\lambda\in \big(2l,2(l+1)\big) \bigcap(l>1)$, we have $j=l-1$ for the single-particle ground states with $j_z=-l+1,-l+2,\cdots,l-1$.
The ground-state is of $2j+1$-fold degeneracy and reads
\begin{align}
	\label{eq:psijz1}
	\psi^{l,1}_{l-1,j_z}(\theta,\phi)=\begin{pmatrix}\sqrt{\frac{(l-j_z)(l-j_z+1)}{2l(2l+1)}}Y_{l,j_z-1}(\theta,\phi) \\
		-\sqrt{\frac{2(l-j_z)(l+j_z)}{2l(2l+1)}}Y_{l,j_z}(\theta,\phi) \\
	\sqrt{\frac{(l+j_z)(l+j_z+1)}{2l(2l+1)}}Y_{l,j_z+1}(\theta,\phi) \end{pmatrix}.
\end{align}

\section{Ground states of the condensates for $\lambda>0$} \label{app:gs}
%%%%%%%%%%%%%%%%%%%%%%%%%%%%%%%%%%%%%%%%%%%%%%%%%%%
\begin{table*}
\renewcommand\arraystretch{1.5}
\centering
\caption{Explicit information of different phases in figure 2 for $\lambda >0$ within different regimes $c_1/c_0$. Here "WF" is short for "wavefunction". "D[a,b,c]" means a diagonal matrix with the diagonal elements $\{a,b,c\}$. Others are the same as those in figure 2. The explicit form of $\alpha$, $\beta$, $\eta_i$, $a$ and $b$ can be found in the context.}
\begin{tabular}{||c||c|c||c|p{1.5cm}<{\centering}|c||c|c|p{2.7cm}<{\centering}|c||}
  \hline
    & mFM$_-(2)$ & P$_-(2)$ & mFM$_-(3)$ & mP$_-(3)$ & P$_-(3)$ & mFM$_-(4)$ & mFM$1_-(4)$ & mP$_-(4)$ & P$_-(4)$ \\
  \hline
   $\frac{c_1}{c_0}$   & $(-\infty,\frac{1}{5})$ & $(\frac{1}{5},\infty)$ & $(-\infty,-\frac{1}{3})$ & $(-\frac{1}{3},\frac{31}{48})$ & $(\frac{31}{48},\infty)$ & $(-\infty,\eta_1 )$ & $(\eta_1,\eta_2)$ & $(\eta_2,\eta_3)$ & $(\eta_3,\infty)$ \\
  \hline
  WF & $\psi^{2,1}_{1, 1}$ & $\psi^{2,1}_{1,0}$ & $\psi^{3,1}_{2,2}$ & $\sqrt{\frac{2}{3}}\psi^{3,1}_{2,-1}+\sqrt{\frac{1}{3}}\psi^{3,1}_{2,2}$ & $\psi^{3,1}_{2,0}$ & $\psi^{4,1}_{3,3}$ & $\psi^{4,1}_{3,2}$ & $\alpha \left [ \psi^{4,1}_{3,-3}+\psi^{4,1}_{3,3} \right ] +\beta \psi^{4,1}_{3,0}$ & $\sqrt{\frac{1}{2}}[\psi^{4,1}_{3,2}+\psi^{4,1}_{3,-2}]$ \\
  \hline
  $|\mathbf{J}|$ & $|\mathbf{J}|=1$ & $|\mathbf{J}|=0$ & $|\mathbf{J}|=2$ & $|\mathbf{J}|=0$ & $|\mathbf{J}|=0$ & $|\mathbf{J}|=3$ & $|\mathbf{J}|=2$ & $|\mathbf{J}|=0$ & $|\mathbf{J}|=0$ \\
  \hline
  $|\mathbf{F}|$ & $|\mathbf{F}|=\frac{1}{2}$ & $|\mathbf{F}|=0$ & $|\mathbf{F}|=\frac{2}{3}$ & $|\mathbf{F}|=0$ & $|\mathbf{F}|=0$ & $|\mathbf{F}|=\frac{3}{4}$ & $|\mathbf{F}|=\frac{1}{2}$ & $|\mathbf{F}|=0$ & $|\mathbf{F}|=0$ \\
  \hline
  $\Delta \mathcal{F}$ & $\frac{\text{D}\{3,3,9\}}{80\pi}$ & $0_{3\times3}$ & $\frac{\text{D}\{25,25,124\}}{630\pi}$ & $\frac{\text{D}\{4,4,4\}}{105\pi}$ & $0_{3\times3}$ & $\frac{\text{D}\{a,a,b\}}{32032\pi}$ & $\frac{\text{D}\{415,415,925\}}{8008\pi}$  & $\frac{\text{D}\{331,331,330\}\alpha^2\beta^2}{1001\pi}$ & $0_{3\times3}$ \\
%  \hline
%  $E_{int}$ & $\frac{21c_0+15c_1}{80\pi}$  &  $\frac{3c_0}{10\pi}$ & $\frac{(199c_0+174c_1)}{(630\pi)}$  & $\frac{(55c_0+24c_1)}{(210\pi)} $ & $\frac{47c_0}{140\pi}$ & $\frac{5(7081c_0+6591c_1)}{96096\pi}$ & $\frac{5(469c_0+351c_1)}{8008\pi}$ & $\frac{(15463346c_0^2-12219324c_0c_1-8328996c_1^2)}{((56465409c_0-69333264c_1)\pi}$ & $\frac{30c_0}{91\pi}$ \\
  \hline
\end{tabular}
\label{WF:informationA}
\end{table*}
%%%%%%%%%%%%%%%%%%%%%%%%%%%%%%%%%%%%%%%%%%%%%%%%%

When $\lambda\in(1,4)>0$, we have $l_0=1$ (see Fig.\ref{fig:angularmomentum_gd}) and the total angular-momentum quantum number $j=\vert l-f\vert=0$.
In this case, spin and orbital-angular-momentum are along opposite directions.
The single-particle ground-state $\psi^{1,1}_{0,0}(\theta,\phi)$ (see Eq.\ref{eq:psi00}) is non-degenerate, which should also be the ground-state for condensates with weak interaction strength.
The system possesses homogeneous density distribution with its spin mean-value $\vert\mathbf{F}\vert=0$ and spin-density $\mathcal{F}_{\alpha}(\theta,\phi)=0~(\alpha=x,y,z)$.
So it inherently belongs to a polar state no matter whether spin-exchange interaction is antiferromagnetic or not.
Core-less vortices with vorticity $n_{\nu}=-1$ and $n_{\nu}=1$ appear in component $f_z=1$ and $f_z=-1$ respectively, which thus also induces pure spin currents in such system.

For larger $\lambda>4$, the explicit ground-states are list in Tab. (\ref{WF:informationA}). 
The table shares the similar pattern as those shown in the main text for $\lambda<0$, which is a reflection of approximated mirror symmetry of the phase diagram around the polar phase when $j=0$.
The explicit expressions appeared in the $j=3$ ($l=4$) case are list as follows
\bea
\alpha^2 &=& \frac{1}{4}-\frac{819c_0}{75212c_0-92352c_1}, ~ \beta=\sqrt{1-2\beta^2},  \nn \\
~ \eta_1 &=& -\frac{1453}{2379} \simeq -0.61,  \nn \\
\eta_2 &=& \frac{(880213-33\sqrt{1848514105})}{2816736} \simeq -0.19, \nn \\
 ~\eta_3 &=& \frac{674}{1443} \simeq 0.47, ~a= 1225, ~ b=8535.  \nn
\eea

\section{Spin vortices in weak interaction $c_0=1$} \label{app:sv}
\begin{figure}[htbp]
	\includegraphics[width=0.5\textwidth]{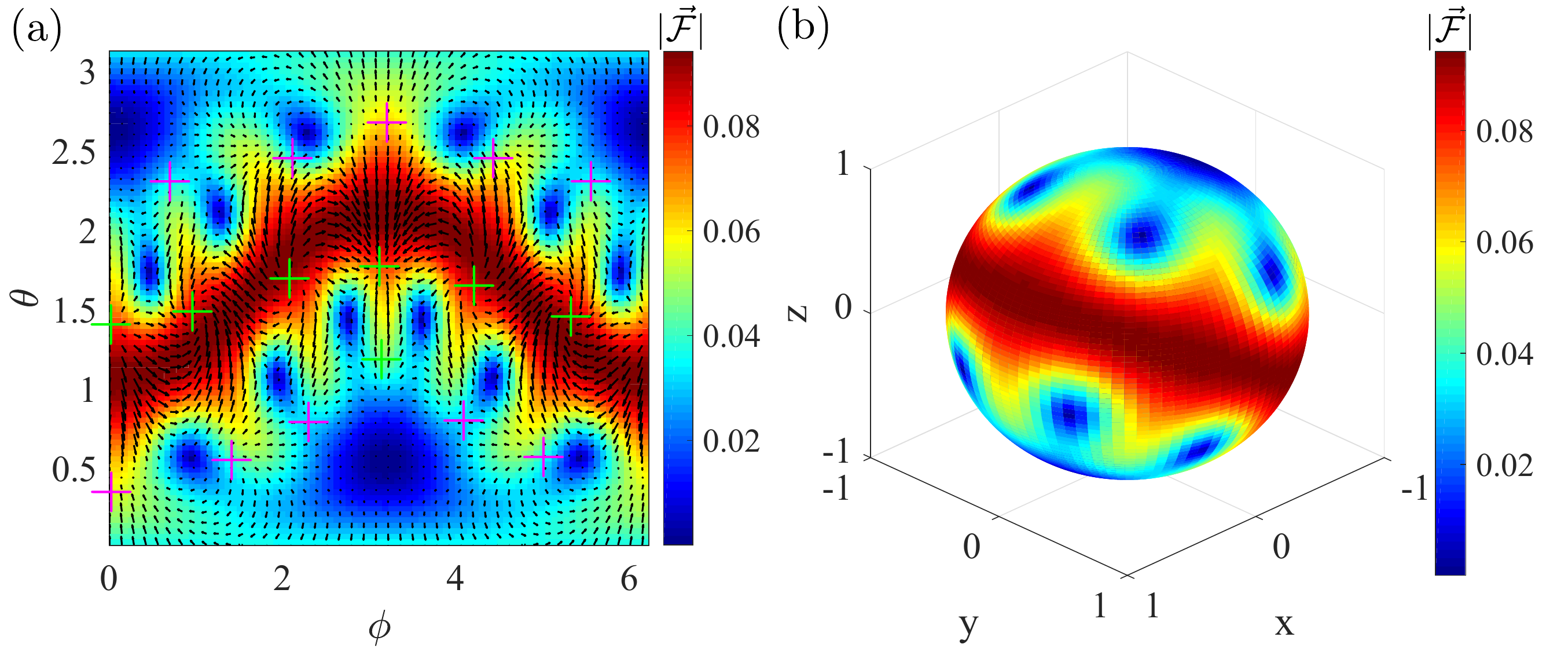}% Here is how to import EPS art
	\caption{Local spin-density vector $\vec{\mathcal{F}}(\theta,\phi)$ in the $\text{mP}_+(2)$ phase with SOAM coupling $\lambda=-5$ and weak interaction $c_0=1$.. Colorful symbols ``+" in (a) represent spin vortices with $Q=1$. There are two kinds of spin vortices with $N_+-N_-=16-14=2$.}\label{fig:topQ1}
\end{figure}

\begin{figure}[htbp]
	\includegraphics[width=0.5\textwidth]{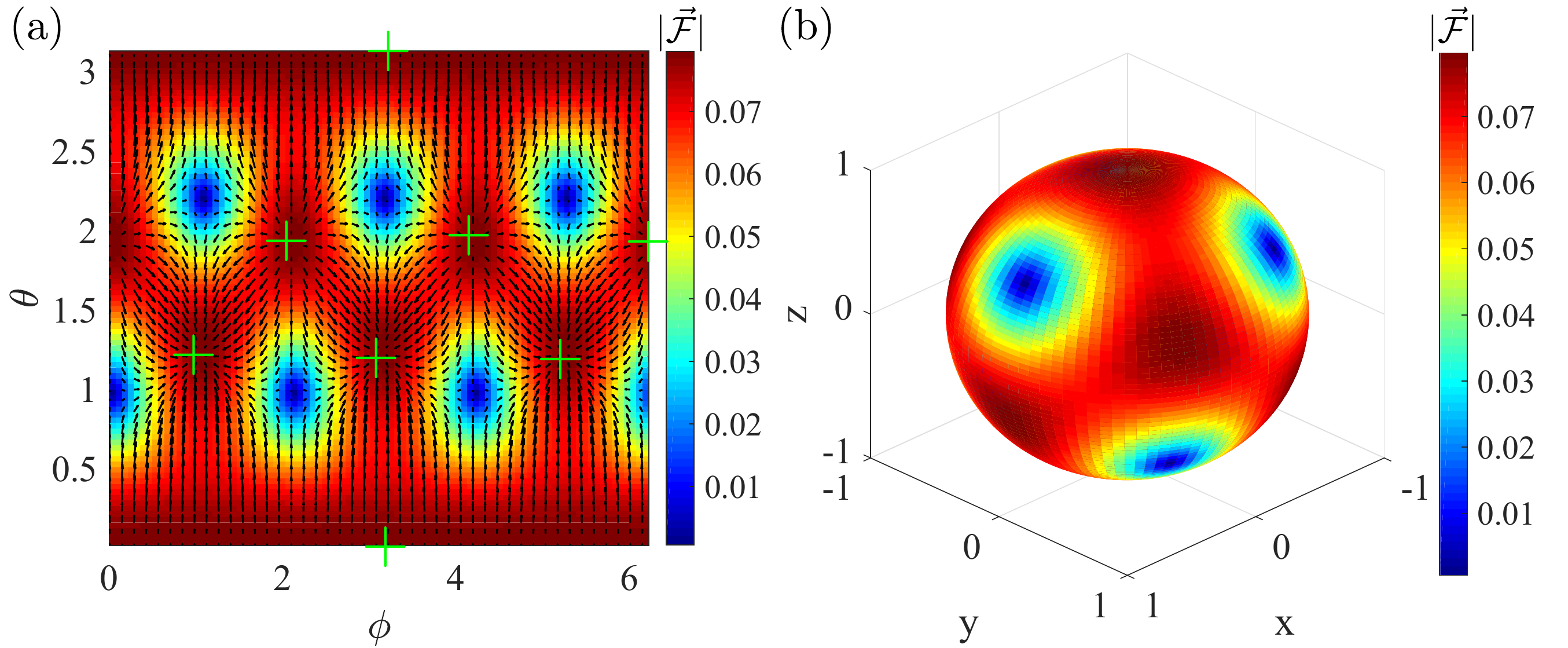}% Here is how to import EPS art
	\caption{Local spin-density vector $\vec{\mathcal{F}}(\theta,\phi)$ in the $\text{mP}_+(1)$ phase with SOAM coupling $\lambda=-3$ and weak interaction $c_0=1$. Colorful symbols ``+" in (a) represent spin vortices with $Q=1$. There are two kinds of spin vortices with $N_+-N_-=8-6=2$.}\label{fig:topQ2}
\end{figure}

\begin{figure}[htbp]
	\includegraphics[width=0.5\textwidth]{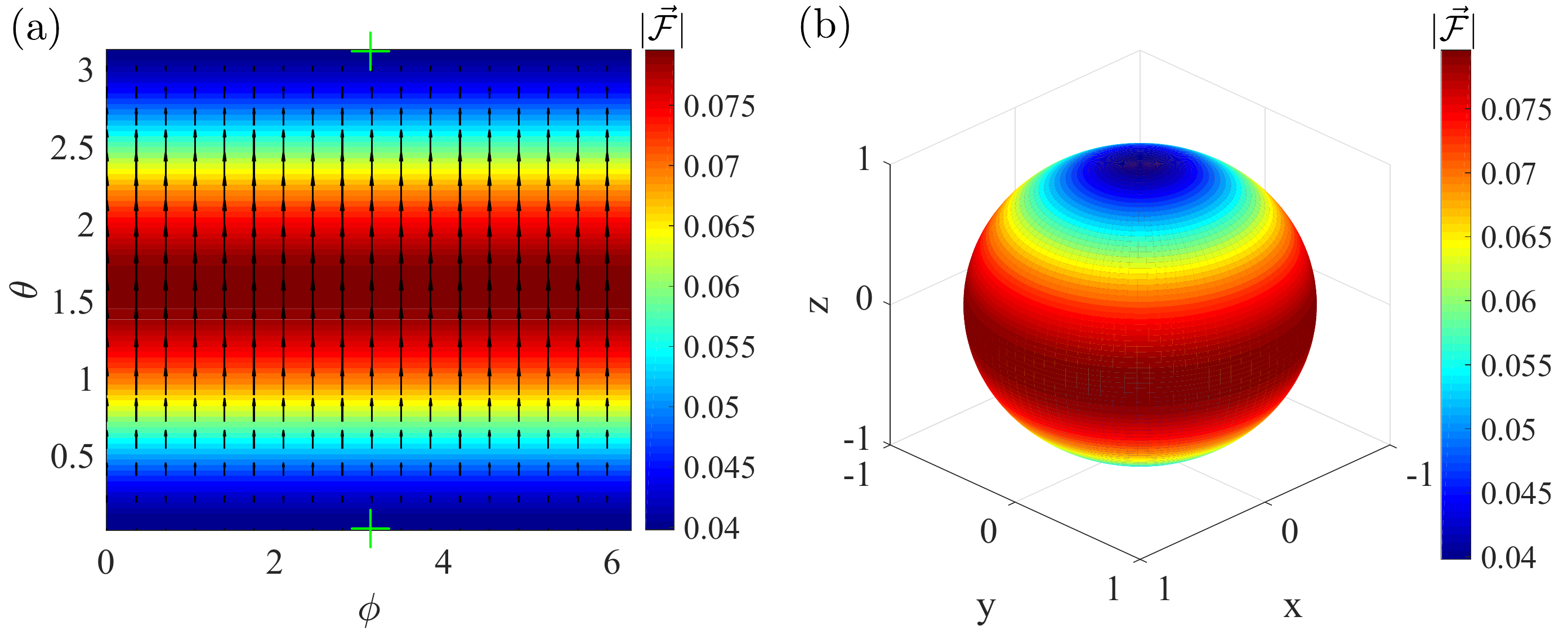}% Here is how to import EPS art
	\caption{Local spin-density vector $\vec{\mathcal{F}}(\theta,\phi)$  in the $\text{mFM}_-(2)$ with SOAM coupling $\lambda=5$ and weak interaction $c_0=1$. Colorful symbols ``+" in (a) represent spin vortices with $Q=1$. There are two spin vortices with $Q=1$.}\label{fig:topQ3}
\end{figure}

\begin{figure}[htbp]
	\includegraphics[width=0.5\textwidth]{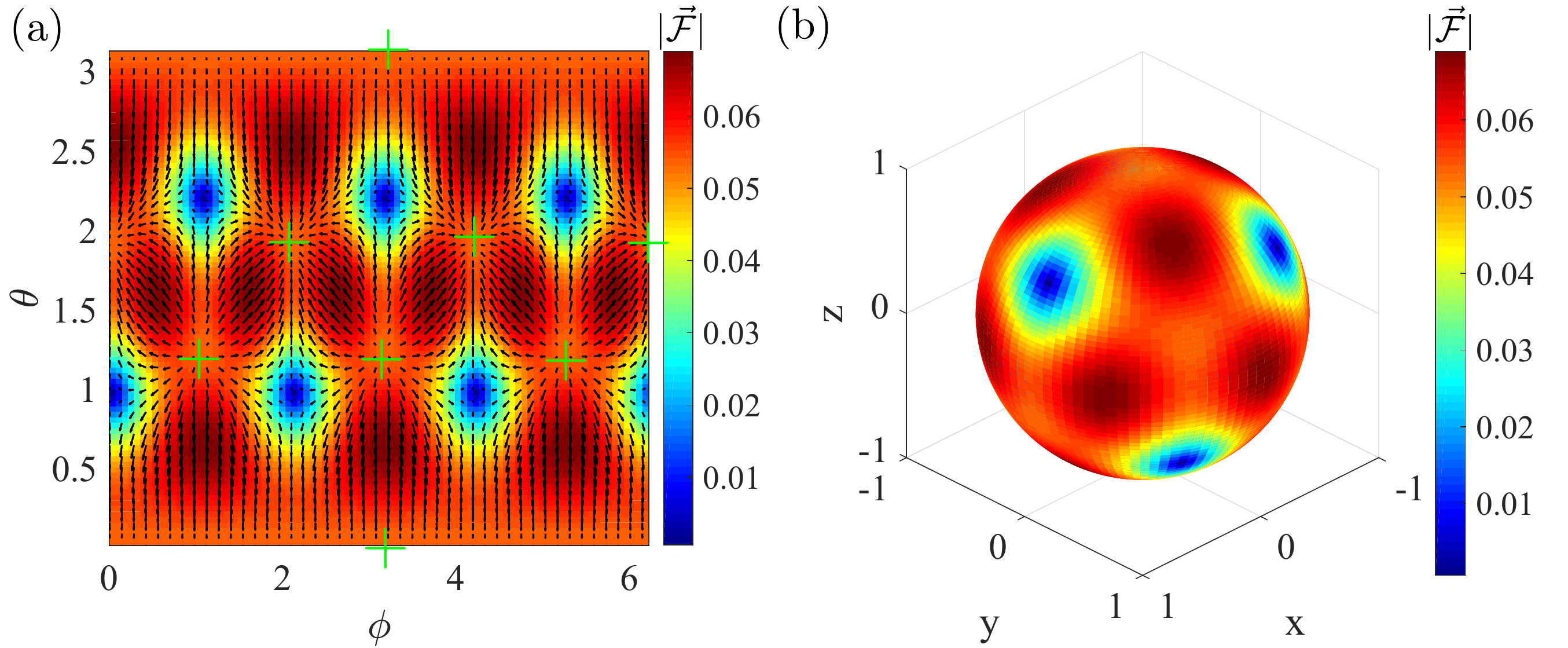}% Here is how to import EPS art
	\caption{Local spin-density $\vec{\mathcal{F}}(\theta,\phi)$ in the $\text{mP}_-(3)$ with SOAM coupling $\lambda=7$ and weak interaction $c_0=1$. Colorful symbols ``+" in (a) represent spin vortices with $Q=1$. There are two kinds of spin vortices with $N_+-N_-=8-6=2$.}\label{fig:topQ4}
\end{figure}

\begin{figure}[htbp]
	\includegraphics[width=0.5\textwidth]{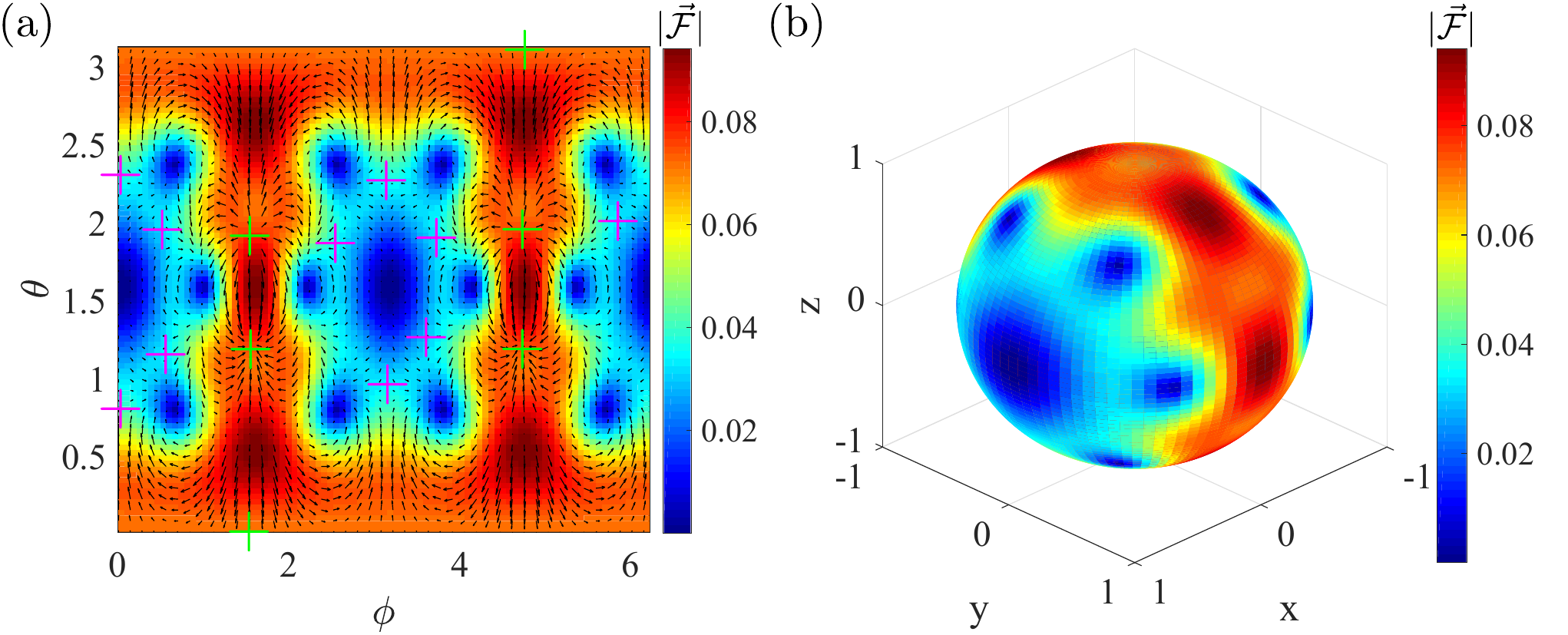}% Here is how to import EPS art
	\caption{Local spin-density $\vec{\mathcal{F}}(\theta,\phi)$  in the $\text{mP}_-(4)$ with SOAM coupling $\lambda=9$ and weak interaction $c_0=1$. Colorful symbols ``+" in (a) represent spin vortices with $Q=1$. There are two kinds of spin vortices with $N_+-N_-=16-14=2$.} \label{fig:topQ5}
\end{figure}

Around $c_1/c_0=0$, the spin-density vector $\vec{\mathcal{F}}(\theta,\phi)=\mathcal{F}_r\mathbf{e}_r+\mathcal{F}_\theta\mathbf{e}_\theta+\mathcal{F}_\phi\mathbf{e}_\phi$ exhibits nontrivial patterns after projecting on the tangent plane of the surface. 
In our case, only spin vortices with Poincar\'{e} index $Q=\pm 1$ exist for $\mathcal{F}_\theta\mathbf{e}_\theta+\mathcal{F}_\phi\mathbf{e}_\phi$ , as shown in Fig.\ref{fig:topQ0}. 
To figure out the distributions of the vector $\vec{\mathcal{F}}(\theta,\phi)$ and  the patterns formed by these spin vortex defects, we list the representative spin-textures in  figures \ref{fig:topQ1}-\ref{fig:topQ5}.
One can see that, in most case with $Q=+1$, we have coreless FM-centered vortices, or mFM-centered coreless vortices. 
While for $Q=-1$, a polar-core spin vortex with $\vec{\mathcal{F}}(\theta,\phi)=0$ at the center is favored.   

When $\lambda\in(4,6)$, an exception occurs and its ground state is meta-ferromagnetic and reads $\psi^{2,1}_{1, 1}$ with maximum mean spin $\mathbf{F}$ (or $\mathbf{J}$, $\mathbf{L}$) and spin fluctuations. The local spin-density vector is $\vec{\mathcal{F}}(\theta,\phi)=-[\cos(\theta)\mathbf{e}_r+2\sin(\theta)\mathbf{e}_{\theta}]/(8\pi)$ as shown in Fig.\ref{fig:topQ3}. So two mFM-centered spin vortices appear in the two poles.

%%%%%%%%%%%%%%%%%%%%%%%%%%%%%%%%%%%%%%%%%%%%%%%%
%\begin{figure}[htbp]
%	\includegraphics[width=0.48\textwidth]{./figs4.eps}% Here is how to import EPS art
%	\caption{%\label{fig:sphe_trap1}
%		Spin vortex defects on the spherical surface in the meta-polar phase, where only defects with two possible Poincar\'{e} index $Q= +1$ and $-1$ are numerically obtained.} \label{fig:topQ0}
%\end{figure}
%%%%%%%%%%%%%%%%%%%%%%%%%%%%%%%%%%%%%%%%%%%%%%%%%%%%%

%%%%%%%%%%%%%%%%%%%%%%%%%%%%%%%%%%%%%%%%%%%%%%%%%%%%%%%%%%%%%

\bibliography{sphBEC-zhou}% Produces the bibliography via BibTeX.

\end{document}